\documentclass[prl,reprint,aps,superscriptaddress,longbibliography]{revtex4-2}

\usepackage[pdftex]{graphicx}

\usepackage{dcolumn}
\usepackage{bm}
\usepackage{amsmath}
\usepackage{amssymb}
\usepackage{ulem}
\usepackage{braket}
\usepackage{bbold}

\usepackage[caption=false]{subfig}
\graphicspath{ {figures/} }

\usepackage{hyperref}
\usepackage[usenames,dvipsnames]{xcolor}
\hypersetup{colorlinks=true,
linkcolor=ForestGreen,
citecolor=RoyalBlue,
filecolor=BrickRed,
urlcolor=RoyalBlue,
filebordercolor={.8 .8 1},
urlbordercolor={.8 .8 0}
}%

\usepackage{tikz}
\usetikzlibrary{positioning}

\newcommand{\<}{\langle}

\renewcommand{\>}{\rangle}
\renewcommand{\(}{\left(}
\renewcommand{\)}{\right)}

\renewcommand{\b}[1]{\mathbf{#1}} 

\def \k{{\mathbf k}}
\def \a{{\mathbf a}}
\def \K{{\mathbf K}}

\newcommand{\moire}{moir\'e }

\begin{document}

\title{Extended symmetric quantum phase in a honeycomb Heisenberg model with sublattice-selective interactions}

\author{Nai Chao Hu}
\affiliation{Department of Physics and Astronomy, Ghent University, Krijgslaan 281, 9000 Gent, Belgium}
\affiliation{Max Planck Institute for Solid State Research, D-70569 Stuttgart, Germany}

\author{Xing-Yu Zhang}
\email{Xingyu.Zhang@Ugent.be}
\affiliation{Department of Physics and Astronomy, Ghent University, Krijgslaan 281, 9000 Gent, Belgium}

\author{Yuchi He}
\affiliation{Department of Physics and Astronomy, Ghent University, Krijgslaan 281, 9000 Gent, Belgium}

\author{Nick Bultinck}
\affiliation{Department of Physics and Astronomy, Ghent University, Krijgslaan 281, 9000 Gent, Belgium}

\begin{abstract}
Motivated by two-dimensional bilayer systems we numerically study an anti-ferromagnetic spin-$1/2$ Heisenberg model on the honeycomb lattice with a nearest-neighbour exchange coupling $J_1$, and a next-nearest-neighbour exchange coupling $J_2'$ for the B sublattice sites only. Using infinite Projected Entangled-Pair States (iPEPS), variational uniform matrix-product states on cylinders, and exact diagonalization, we find an extended symmetric regime $0.4\lesssim J'_2/J_1\lesssim0.6$, in which local observables show no magnetic, valence-bond, or chiral spin order. At $J'_2/J_1=0.5$, the PEPS correlation length grows systematically with bond dimension, and an inverse-correlation-length extrapolation favors a vanishing limit. We also find that various observables scale algebraically with the finite bond-dimension-induced correlation length, which points to a gapless spin liquid ground state. We propose a $\mathbb{Z}_2$ Dirac spin liquid parton state, with Dirac points that are protected by translation, time-reversal and three-fold rotation symmetry, as a promising candidate state to explain the numerical results. We also discuss the possibility that the symmetric ground state is a featureless, short-range entangled state with a small gap.
\end{abstract}

\maketitle

\paragraph*{Introduction ---}
Frustration is one of the principal routes by which quantum fluctuations destabilize conventional symmetry-breaking orders and promote exotic phases of matter. In Heisenberg models on the square, honeycomb, and triangular lattices, competing interactions can produce intermediate quantum-disordered regimes, including quantum spin liquids (QSLs) and valence-bond solids (VBSs) \cite{Anderson1973,Wen2002,Balents2010}. These regimes, however, typically occupy limited windows between ordered phases, making them difficult to identify unambiguously in numerical studies and to access without careful tuning in experiments. Finding simple, physically motivated forms of frustration that stabilize quantum disorder over a broad range of parameters therefore remains an important goal.

\begin{figure}[t]
    \centering
    \includegraphics[width=\linewidth]{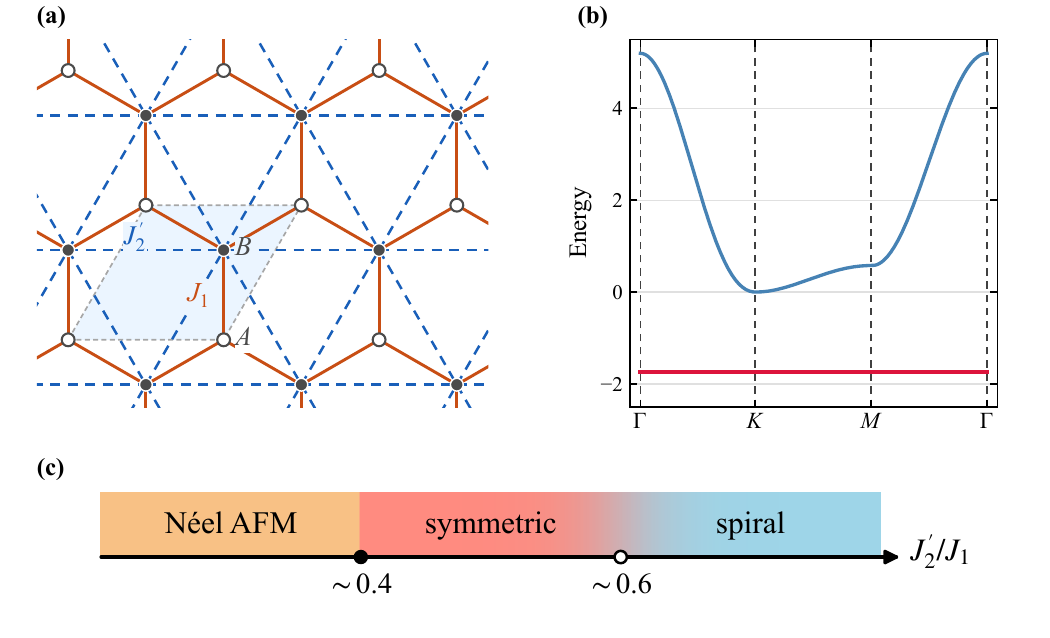}
    \caption{(a) Honeycomb lattice with nearest-neighbor $J_1$ exchange on the solid orange bonds and next-nearest-neighbor $J'_2$ exchange on the dashed blue bonds connecting only $B$-sublattice sites. The shaded region denotes a primitive unit cell. (b) Spectrum of the corresponding Fourier transformed exchange coupling matrix at $J'_2/J_1=1/\sqrt{3}$, where the lower band is exactly flat. (c) Schematic quantum phase diagram inferred from the numerical results. N\'eel order gives way near $J'_2/J_1\simeq0.4$ to an extended symmetric quantum-disordered regime; beyond $J'_2/J_1\simeq0.6$, finite-momentum spiral correlations become dominant, for which we loosely use the label ``spiral''.}
    \label{fig:hc}
\end{figure}

In this Letter, we study an anti-ferromagnetic honeycomb-lattice model in which a next-nearest-neighbor (NNN) exchange interaction acts on only one sublattice [Fig.~\ref{fig:hc}(a)]. Such an interaction arises naturally in bilayer systems, where each sublattice corresponds to a triangular lattice on one of the two layers, a setup which is realized in moir\'e transition metal dichalcogenide (TMD) systems~\cite{moireKondo,zhao2023gate}. The sublattice-selective interaction places the spin model in a maximally frustrated regime. To understand this from a classical perspective, we note that at $J_2'/J_1=1/\sqrt{3}\approx0.577$, with $J_1$ the nearest-neighbour (NN) and $J_2'$ the sublattice-selective NNN exchange coupling, the lowest band of the Fourier-transformed exchange matrix is exactly flat [Fig.~\ref{fig:hc}(b)], just as in the kagome anti-ferromagnet. The flat band can be understood as a Tasaki-type flat-band construction \cite{PhysRevLett.69.1608, tanaka2020extension}, or as being spanned by the compact localized states~\cite{PhysRevB.78.125104} with local operators $A^{\dagger}_{\b R} = \frac{1}{\sqrt{6}}(\sqrt{3}c^{\dagger}_{B,\b R}-\sum_{m=1}^3c^{\dagger}_{A,\b R+\bm \delta_m})$ (in the fermionic language), with $\bm \delta_m$ the 3 NN vectors on the $B$ sublattice site at $\b R$. The exactly flat band implies that a classical Luttinger-Tisza-Lyons-Kaplan analysis \cite{luttinger-tisza, lyons-kaplan} cannot select a unique wave vector for magnetic order in this system, and there is an extensive ground state degeneracy -- an ideal scenario for quantum fluctuations to determine the low-temperature physics.

To address the full quantum problem, we combine infinite projected entangled-pair states (iPEPS), symmetric variational uniform matrix-product states (VUMPS) on infinite cylinders, and exact diagonalization (ED) to obtain the phase diagram in Fig.~\ref{fig:hc}(c). We find an extended symmetric regime, approximately $0.4\lesssim J_2'/J_1\lesssim0.6$, encompassing the flat-band point. Local observables and correlation functions show no magnetic, valence-bond, or chiral order, while transfer-matrix and spin-flux-insertion diagnostics point to the absence of gapped topological order. Although the correlation lengths remain only a few lattice spacings, the finite-correlation-length scaling of the energy-density correction and the bond dimension dependence of the correlation length point to a $z=1$ gapless state. We propose a $\mathbb{Z}_2$ Dirac spin liquid parton state which is consistent with the numerical results, including previous simulations on the conventional spin-$1/2$ $J_1$-$J_2$ model on the honeycomb lattice~\cite{Ferrari2017}. One particularly appealing feature of this parton state is that it does not require inversion symmetry, which is broken by the sublattice-selective interaction, to protect the Dirac points. Finally, we note that on a honeycomb lattice of spin-$1/2$'s there is no Lieb-Schultz-Mattis (LSM) obstruction~\cite{LSM1961, Oshikawa00, Hastings2004,watanabe2015filling, PhysRevLett.119.127202} to a featureless, short-range entangled state (SRES) \cite{kimchiPNAS, Jian2016featureless,kim2016featureless}. Our numerics cannot rule out such a SRES with a small gap. We also discuss a realization of the spin model considered here in moir\'e materials.

\begin{figure*}
    \centering
    \includegraphics[width=\textwidth]{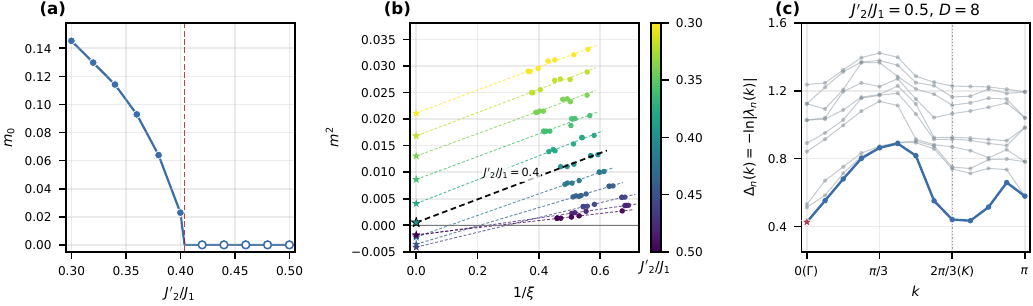}
    \caption{Magnetization extrapolations and transfer-matrix spectrum from merge-optimized iPEPS calculations using $D=6,7,8,10$ tensors with environment bond dimensions up to $\chi=1024$. (a) Extrapolated magnetization extracted from Eq.~\eqref{xisc}; open symbols denote negative $m_0^2$ intercepts plotted as $m_0=0$, and the dashed line marks the fitted zero crossing. (b) Finite-$\xi$ magnetization data and linear extrapolations. Colors encode $J'_2/J_1$, stars mark the intercepts at $1/\xi=0$, and the black dashed line highlights $J'_2/J_1=0.4$. (c) Transfer-matrix excitation spectrum at $J'_2/J_1=0.5$ and $D=8$. The lowest branch is highlighted in blue, gray curves show the higher branches, the red star marks the global minimum, and the vertical dotted line marks $K$ at $k=2\pi/3$.}
    \label{fig:peps-mag}
\end{figure*}

\paragraph*{Model and classical limit ---} The spin-$1/2$ honeycomb lattice Hamiltonian we consider is (see Fig.~\ref{fig:hc}):
\begin{align}
    H = J_1\sum_{\<{\bm{r},\bm{r'}}\>}\bm{S}_{\bm{r}}\cdot \bm{S}_{\bm{r'}} + J'_2\sum_{\<{\<{\bm{r},\bm{r'}}\>}\>,\bm{r}\in B}\bm{S}_{\bm{r}}\cdot \bm{S}_{\bm{r'}}, \label{eq:2d}
\end{align}
where $\<\cdot\>$ and $\<\<\cdot\>\>$ represent NN and NNN bonds. The NNN interaction is only between B sublattice sites. Classically, the model interpolates between the honeycomb N\'eel state at small $J'_2$ and a noncollinear ferrimagnet at large $J'_2$, where the $B$ sublattice forms a canted $120^\circ$ pattern while the $A$ spins are ferromagnetic, as shown in Fig.~\ref{fig:cl-spin}. The flat-band point mentioned in the introduction lies between these limits and marks a maximally frustrated regime. Details of the classical variational state and its energy are given in~\cite{SM}.

\paragraph*{Magnetization ---}
For $J'_2/J_1 \lesssim 0.4$, the ground state retains the N\'eel order of the nearest-neighbor honeycomb anti-ferromagnet. In our iPEPS simulations, we calculate the local magnetization for different values of $J_2'/J_1$ and for different bond dimensions $D$. The PEPS is contracted by finding the fixed point of the transfer matrix, approximated as a matrix product state with environment bond dimension $\chi$ (the fixed point is obtained with the VUMPS algorithm). For our results we have used values of $\chi$ up to $1024$ for optimization. For further details on the iPEPS simulations we refer to \cite{SM}.

The PEPS bond dimension $D$ induces a finite correlation length $\xi$. In the N\'eel phase we find that the squared magnetization $m^2$ follows the typical finite-correlation-length scaling \cite{PhysRevLett.129.200601}:
\begin{equation}
    m^2(\xi) \approx m_0^2 + a/\xi \label{xisc}
\end{equation}
In Fig.~\ref{fig:peps-mag}(b) we show the numerical results for $m^2(\xi)$. These results were obtained with a translationally invariant PEPS ansatz that has an $A$ and $B$ site blocked together in a single tensor \cite{SM}. The extrapolated magnetic order parameter $m_0$ decreases rapidly with increasing $J'_2$ and vanishes near $J'_2/J_1\simeq 0.4$, consistent with the breakdown point of the Anderson tower of states in ED on a 24-site cluster (see \cite{SM}). Between $J'_2/J_1\simeq0.4$ and $0.6$, we find that the interpolation in Eq. \eqref{xisc} produces a negative $m_0^2$ value. This signals a breakdown of the relation in Eq. \eqref{xisc}, which points to the absence of N\'eel order.

\paragraph*{VBS and time-reversal breaking order parameters ---} We now test for other types of symmetry breaking in the range $0.4\lesssim J'_2/J_1 \lesssim 0.6$. One obvious type of order we need to rule out is a VBS state, in which rotation and/or translation symmetry is spontaneously broken by a bond modulation pattern. A plaquette VBS is commonly found in the standard $J_1$-$J_2$-$J_3$ model on the honeycomb lattice \cite{PhysRevB.84.024406}. We check for plaquette VBS order at $J'_2/J_1=0.5$ using several larger-unit-cell PEPS constructions initialized with plaquette VBS states obtained from the $J_1$-$J_2$-$J_3$ model. Their energies nearly saturate with increasing $D$, and none improves upon the best translationally invariant PEPS energy. We also optimize a larger unit cell initialized by tiling the optimized uniform tensor and find no lower-energy PVB state~\cite{SM}. Columnar VBS order can be captured by the translationally invariant ansatz, but does not occur in our simulations.

\begin{figure}
    \centering
    \includegraphics[width=\linewidth]{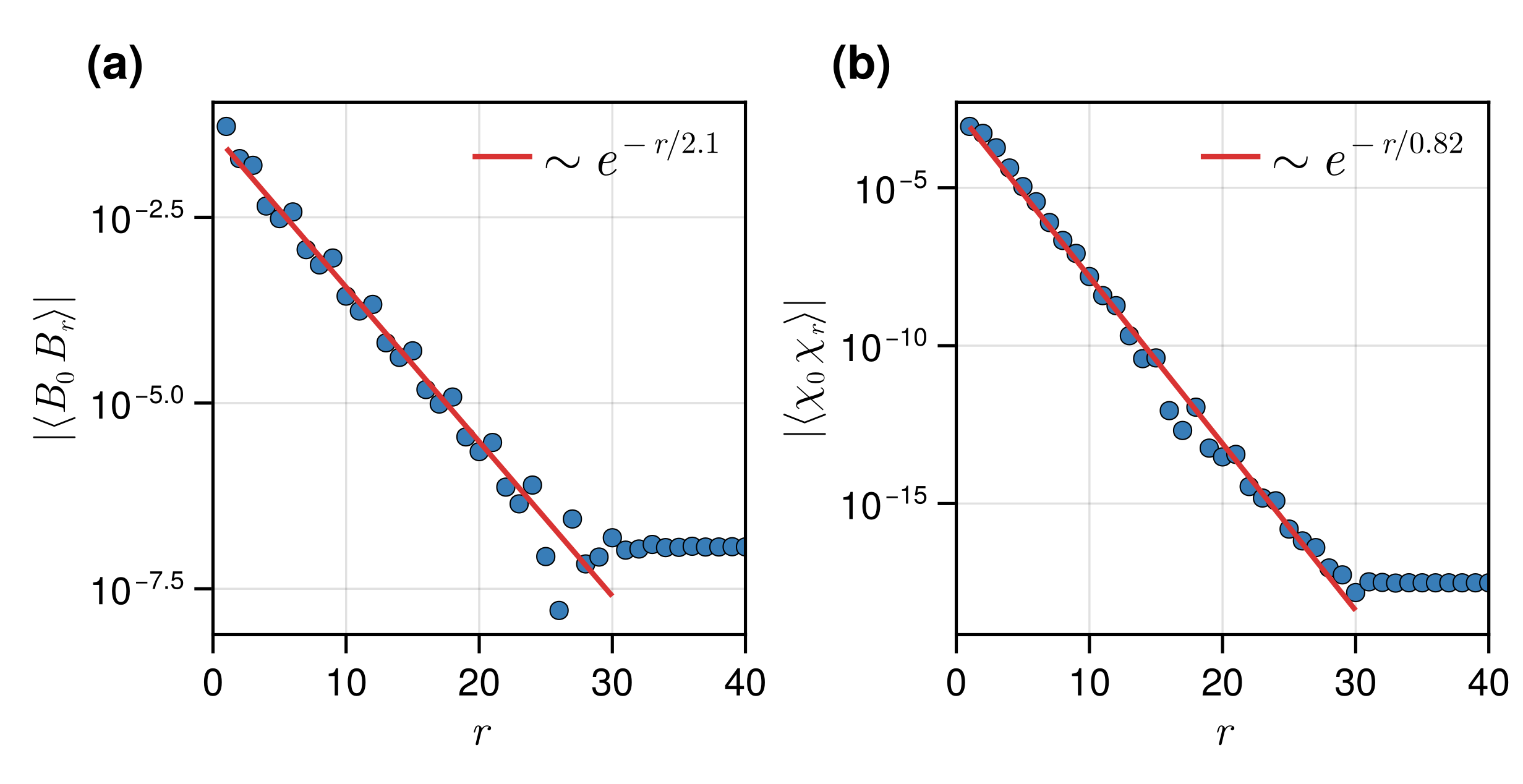}
    \caption{Bond-bond and scalar-chirality correlations from YC12 cylinder VUMPS at $J_2'=0.5J_1$ and $D=6400$, where we use the $C_3$-breaking order parameter $B_r=\mathbf{S}_1\cdot\mathbf{S}_2 - \mathbf{S}_2\cdot\mathbf{S}_3$ and chirality order parameter $\chi_S=\mathbf{S}_1\cdot(\mathbf{S}_2\times\mathbf{S}_3)$. Here $r$ labels distance along the infinite direction of the cylinder. The plotted absolute values decay exponentially, with fitted correlation lengths $\xi_B\simeq 2.1$ and $\xi_\chi\simeq 0.82$ (not converged in bond dimension~\cite{SM}), giving no evidence for long-range VBS or chiral order.}
    \label{fig:mps-corr}
\end{figure}

The absence of VBS order is further supported by cylinder VUMPS calculations with exact SU(2) symmetry: for sufficiently wide cylinders, bond inhomogeneities quickly vanish as the bond dimension increases, despite the rotational-symmetry breaking imposed by the cylinder geometry~\cite{SM}. 
Fig.~\ref{fig:mps-corr}(a) further shows that the plaquette and columnar VBS order parameter correlation function $\<(\mathbf{S}_1\cdot\mathbf{S}_2- \mathbf{S}_2\cdot\mathbf{S}_3)_0(\mathbf{S}_1\cdot\mathbf{S}_2- \mathbf{S}_2\cdot\mathbf{S}_3)_r\>$ decays exponentially to a small value $\sim 10^{-7}$ along the infinite direction of the cylinder. The correlation length in this channel is the same as the largest correlation length obtained from the MPS transfer-matrix spectrum.  We caution however that the correlation lengths for this circumference are not converged in bond dimension~\cite{SM}.

We also consider the possibility of time-reversal symmetry breaking. In the iPEPS simulations, allowing complex tensor entries or restricting the tensors to be strictly real produces identical energies and  observables, showing the absence of time-reversal symmetry breaking. In cylinder VUMPS [see Fig.~\ref{fig:mps-corr}(b)], we check that the expectation value of the chirality operator $\chi_S=\mathbf{S}_1\cdot(\mathbf{S}_2\times\mathbf{S}_3)$ is always zero within numerical precision. Fig.~\ref{fig:mps-corr}(b) further shows that the chirality correlations are short-ranged and much weaker than the VBS correlations.

\paragraph*{Onset of finite wave-vector correlations ---} The upper boundary $J'_2/J_1\simeq0.6$ of the quantum disordered phase is estimated by the transfer-matrix spectrum of the translationally-invariant PEPS developing a second largest eigenvalue near momentum $k\sim 2\pi/3$ (see Figs.~\ref{fig:peps-mag}(c) and \ref{fig:peps-tm-spectrum}). This signals the onset of a parameter region where the dominant correlations occur at a non-zero wave-vector. This is consistent with the finite-momentum peaks found in the cylinder VUMPS spin structure factors \cite{SM}. The same coupling range exhibits a dense set of competing low-energy momentum sectors and evolving sublattice-resolved structure-factor peaks in ED on the 24-site cluster \cite{SM}. All these findings point to a regime with spiral spin correlations. Since both ED and cylinder VUMPS show that the two sublattices exhibit different dominant spiral wave vectors~\cite{SM}, the exact nature of this phase remains difficult to determine with our current methods and is left for future work.

\begin{figure}[!t]
    \centering
    \includegraphics[width=\columnwidth]{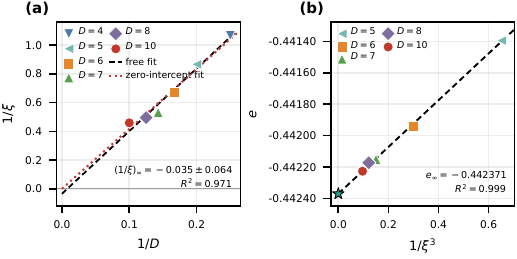}
    \caption{iPEPS correlation-length and energy scaling at $J'_2/J_1=0.5$. (a) Finite-$D$ scaling of $1/\xi$ using $D=4,5,6,7,8,10$ and the largest available converged environment at each $D$. The black dashed line is an unconstrained linear fit, which yields $(1/\xi)_\infty=-0.035\pm0.064$, indistinguishable from zero; the red dotted line is constrained to pass through the origin. (b) Energy density versus $1/\xi^3$ for the five $D=5,6,7,8,10$ points used in the magnetization fit. The dashed line is a linear fit, with the star marking $e_\infty=-0.442371$.}
    \label{fig:peps-xi-tm}
\end{figure}

\paragraph*{Absence of gapped topological order ---}
We now also rule out that the symmetric quantum phase at $J_2'/J_1 = 0.5$ has non-trivial gapped topological order. The first evidence is that the transfer matrix of the converged iPEPS has a non-degenerate largest eigenvalue. The uniqueness of the transfer matrix fixed point rules out gapped topological order \cite{schuch2010peps, PhysRevLett.111.090501, haegeman2015shadows}.

In cylinder VUMPS simulations we have performed adiabatic spin flux insertion. For the spin model in Eq.~\eqref{eq:2d}, we modify the spin operators $S^+_{x,y}S^-_{x,y+1} \rightarrow S^+_{x,y}S^-_{x,y+1}e^{i\Phi/L_y}$ to include a flux $\Phi$. This reduces the spin SU(2) symmetry of the MPS tensors to U(1). By slowly varying the flux from 0 to $2\pi$, we can adiabatically change the Hamiltonian $H \rightarrow H(\Phi)$. After a $2\pi$ flux cycle, $H(\Phi = 2\pi)$ is related to $H(\Phi = 0)$ by a gauge transformation: $UH(2\pi)U^{-1} = H(0)$. We adjust the final state back to the same gauge as the initial state, and find $|\<\Psi(0)|U|\Psi(2\pi)\>|=1$ within numerical precision for every cylinder width we studied. From this we conclude that there are no topologically distinct ground states with an energy splitting that is exponentially small in $L_y$~\cite{He2017}.

\paragraph*{Evidence for a gapless spin liquid ---} At $J'_2/J_1=0.5$, the PEPS correlation length obtained from the largest available converged environment at each tensor bond dimension increases monotonically from $\xi=0.94$ at $D=4$ to $\xi=2.18$ at $D=10$. Fig.~\ref{fig:peps-xi-tm}(a) shows the corresponding inverse-correlation-length extrapolation using $D=4,5,6,7,8,10$. The unconstrained linear fit gives $(1/\xi)_\infty=-0.035\pm0.064$ with $R^2=0.971$, i.e. an extrapolated inverse correlation length indistinguishable from zero. This is compatible with a gapless state, but we caution that the intercept is sensitive to which bond dimensions enter the fit, since $1/\xi$ falls by $0.14$ between $D=6$ and $D=7$ but only by $0.04$ between $D=8$ and $D=10$, so a straight line through the large-$D$ end alone extrapolates to a positive value. It is possible that for the largest $D$ values one would need to further increase $\chi$ to obtain a converged correlation length, but this was not possible with our numerical resources. In general, we observe the need to use much larger $\chi$ than the empirical relation $\chi \sim D^2$ for reliable energy optimization. For example, for $D=8$ and $10$ we used a maximal value of $\chi=1024$. This also points to the state being gapless. 

Additional evidence supporting a gapless spin liquid is provided by the iPEPS energy density, which depends on the correlation length as $e(\xi)-e(\infty) \sim 1/\xi^\alpha$ (see Fig.~\ref{fig:peps-xi-tm}(b)). $\alpha = 3$ is the expected scaling for a Lorentz-invariant ($z=1$) gapless state~\cite{neuberger1989finite, schulz1996magnetic}. However, $\alpha = 2, 4$ can also fit the accessible correlation length range reasonably well, leaving the exponent not uniquely resolved.

\begin{figure}
    \centering
    \includegraphics[scale=0.35]{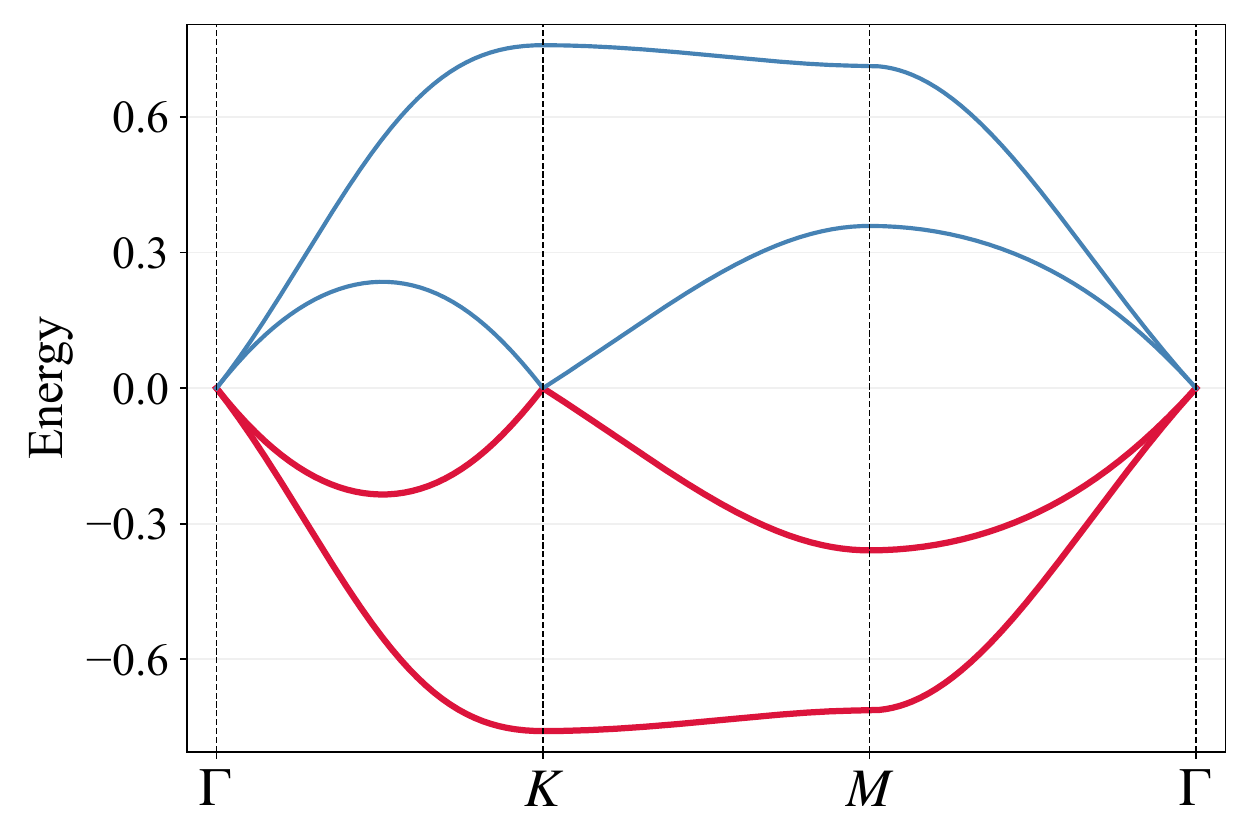}
    \caption{Spinon spectrum of the Bogoliubov–de Gennes Hamiltonian Eq.~\eqref{eq:hz2}, with $\Delta_A(\k)=0$ and $\Delta_B(\k)$ finite (see text). Compared to the gapless state in Ref.~\cite{Ferrari2017}, the sublattice asymmetry lifts the double degeneracy along the high-symmetry line $M\rightarrow\Gamma$, leaving the Dirac points untouched.}
    \label{fig:z2dirac}
\end{figure}

\paragraph*{$\mathbb{Z}_2$ Dirac spin liquid parton state ---} We propose the parton mean-field Hamiltonian $H_{mf} = \sum_\k \psi^\dagger_\k H(\k) \psi_\k$ as a promising candidate for the gapless spin liquid, where $\psi_\k = (c_{A,\k,\uparrow},c_{B,\k,\uparrow},c^\dagger_{-\k,A,\downarrow},c^\dagger_{-\k,B,\downarrow})^T$, and
\begin{equation}
H(\k) = \left(\begin{matrix} h(\k-\K) & \Delta(\k) \\ \Delta^* (\k)& -h(\k+\K) \end{matrix}\right)\label{eq:hz2}
\end{equation}
Here, $h(\k)$ is the NN hopping graphene Hamiltonian: $h(\k) = \text{Re}f(\k) \sigma^x + \text{Im}f(\k) \sigma^y$ with $f(\k)=\sum_{m=1}^3 e^{i\boldsymbol{\delta}_m\cdot \k}$.
The Dirac points of $h(\k)$ are at $\pm\K = \pm(\frac{4\pi}{3},0)$. The pairing term is given by
\begin{equation}
\Delta(\k) = \left(\begin{matrix} \Delta_{A}(\k) & 0 \\ 0 & \Delta_B(\k) \end{matrix}\right)
\end{equation}
with $\Delta_A(\k)=0$ and $\Delta_B(\k) = \cos(\k\cdot \a_1) + e^{2\pi i/3} \cos(\k\cdot \a_2) + e^{-2\pi i/3} \cos(\k\cdot\a_3)$. The vector $\a_1$ is the Bravais lattice vector in the $x$-direction, and $\a_2,\a_3$ are related to $\a_1$ by respectively a $2\pi/6$ and $4\pi/6$ rotation. In Ref.~\cite{Ferrari2017} it was shown that the Gutzwiller-projected mean-field state with $\Delta_A(\k)=\Delta_B(\k)$ (No. 18 in the projective symmetry group classification of Ref.~\cite{Lu2011}) gives very competitive variational energies for the standard $J_1-J_2$ honeycomb model in a region of the phase diagram between the N\'eel phase and the plaquette VBS. An interesting property of this spin liquid is that the Dirac points are robust even after breaking the inversion symmetry -- in Fig.~\ref{fig:z2dirac} we plot the mean-field spectrum with the inversion-breaking pairing $\Delta_A(\k)=0$, which clearly still shows the Dirac spectrum. To understand this, we note that the time-reversal symmetry acts as $T=\sigma^z \mu^x K$, where $K$ is complex conjugation and $\mu^i$ are the Pauli matrices in Nambu space. Combined with the particle-hole symmetry $C = \mu^y K$, the time-reversal symmetry gives rise to a chiral symmetry $CT=\sigma^z \mu^z$: $\{H(\k),\sigma^z\mu^z\} = 0$. Together with translation and three-fold rotation symmetry, one can show that the chiral symmetry protects the eight zero modes in this model~\cite{Koshino2014}. This is because for the four zero modes $|\chi^\Gamma_m\rangle$ at the $\Gamma$ point one finds that $C_{3z}|\chi^\Gamma_m\rangle = e^{\eta_m \pi i/3}|\chi^\Gamma_m\rangle$ \footnote{Note that $(C_{3z})^3=(-1)^N$, with $N$ the fermion number.} and $\sigma^z\mu^z|\chi^\Gamma_m\rangle = \eta_m |\chi^\Gamma_m\rangle$, with $\eta_m = +1$ for two of the zero modes, and $\eta_m=-1$ for the other two zero modes. For the four zero modes at $\pm \K$ one finds $C_{3z}|\chi^{\pm \K}_m\rangle = e^{-\eta_m \pi i/3}|\chi^{\pm \K}_m\rangle$ and $\sigma^z\mu^z|\chi^{\pm \K}_m\rangle = \eta_m |\chi^{\pm \K}_m\rangle$. This shows that the zero modes cannot be removed without breaking either the chiral symmetry (and hence the time-reversal symmetry), the three-fold rotation or the translation symmetry.

\paragraph*{Moir\'e realization ---}
The exchange pattern considered in this work has a natural realization in \moire TMD heterostructures. In the Kondo-lattice setting proposed in Ref.~\cite{moireKondo} and experimentally realized in Ref.~\cite{zhao2023gate}, the two active layers experience different \moire potentials, so one layer is localized while the other remains itinerant. Reducing the twist angle enhances the \moire potential and can localize both active layers, producing a spin model with strongly layer-dependent hopping amplitudes. Identifying these hoppings with the two triangular sublattices gives a direct route to the sublattice-asymmetric exchange in Eq.~\eqref{eq:2d}.
The self-consistent Hartree-Fock parameters of Ref.~\cite{moireKondo} with $V_{mf}=30$~meV already give a sizable asymmetry. At a twist angle $\theta=3.5^\circ$, the hopping parameters in the two layers are $t_f=2.72$~meV and $t_c=5.98$~meV. Using the notation of Ref.~\cite{moireKondo},
a strong-coupling estimate for the NNN exchange couplings in the two layers gives $J_2''\simeq 4t_f^2/U$ and $J_2'\simeq 4t_c^2/U_c$, so $J_2'/J_2''\gtrsim (t_c/t_f)^2\approx 4.8$. The asymmetry increases at smaller twist angles, suggesting that the exchange pattern required by the flat-band honeycomb model is realistic in this platform.

\paragraph*{Discussion ---} We have introduced a simple honeycomb spin model in which sublattice-selective interactions produce an extended symmetric quantum phase. Combining large-scale tensor-network simulations with exact diagonalization, we find that Néel order disappears near $J'_2/J_1\simeq0.4$. Within an extended regime, $0.4\lesssim J_2'/J_1\lesssim0.6$, the uniform symmetric state has a lower variational energy than various competing plaquette states we considered.

We proposed two possible scenarios for this symmetric quantum phase: One is a gapless $\mathbb{Z}_2$ Dirac spin liquid, and the other is a featureless SRES. If the SRES scenario is realized, this would be -- to the best of our knowledge -- the first example of a featureless paramagnet in the ground state phase diagram of a spin-$1/2$ Heisenberg model on the honeycomb lattice. In Ref.~\cite{kim2016featureless} a $D=4$ trial state PEPS for such a featureless SRES was constructed, which respects SU(2) spin rotation and all lattice symmetries. If we re-evaluate that trial state with modern VUMPS contraction methods, we find that it has a correlation length which grows polynomially with environment bond dimension $\chi$, with no sign of saturation. At $\chi=1024$ we find $\xi \sim 60$~\cite{SM}. This further illustrates that constructing a featureless paramagnet on the honeycomb lattice is a subtle task, and might potentially require more entanglement than was originally anticipated.

\paragraph{Acknowledgments ---}
We acknowledge the EuroHPC Joint Undertaking for awarding project EHPC-DEV-2026D04-006 access to MareNostrum~5, hosted by the Barcelona Supercomputing Center (BSC), Spain.
This research was supported by the European Research Council under the European Union Horizon 2020 Research and Innovation Programme via Grant Agreement No. 101076597-SIESS (N.C.H. and N.B.).
X.Y.Z. and Y.H. acknowledge support from the European Union through the European Research Council (ERC, GaMaTeN, Grant No. 101125822). Views and opinions expressed are, however, those of the authors only and do not necessarily reflect those of the European Union or the European Research Council Executive Agency. Neither the European Union nor the granting authority can be held responsible for them.

\paragraph*{Data availability ---} The numerical data prepared for release are deposited in a dedicated repository~\cite{HoneycombFlatData}, which will be made publicly accessible upon the appearance of the arXiv preprint.

\bibliography{ref}

\clearpage
\onecolumngrid

\begin{center}
\textbf{\large Supplemental Material for ``Extended symmetric quantum phase in a honeycomb Heisenberg model with sublattice-selective interactions''}
\end{center}

\setcounter{equation}{0}
\setcounter{figure}{0}
\setcounter{table}{0}
\setcounter{page}{1}
\setcounter{section}{0}
\setcounter{secnumdepth}{1}
\makeatletter
\renewcommand{\theequation}{S\arabic{equation}}
\renewcommand{\thefigure}{S\arabic{figure}}
\renewcommand{\thesection}{S\arabic{section}}

This supplemental material includes details on: (i) the classical variational state of the $J_1-J_2'$ model; (ii--iii) the iPEPS optimization, Plaquette-VBS variational comparison, correlation-length scaling, and transfer-matrix calculations; (iv) symmetric MPS calculations on cylinders; (v) exact diagonalization; and (vi) the VUMPS contraction of the trial state PEPS for a featureless state on the honeycomb lattice.

\tableofcontents

\section{Classical variational state}

\begin{figure}[b]
    \centering
    \includegraphics[width=\linewidth]{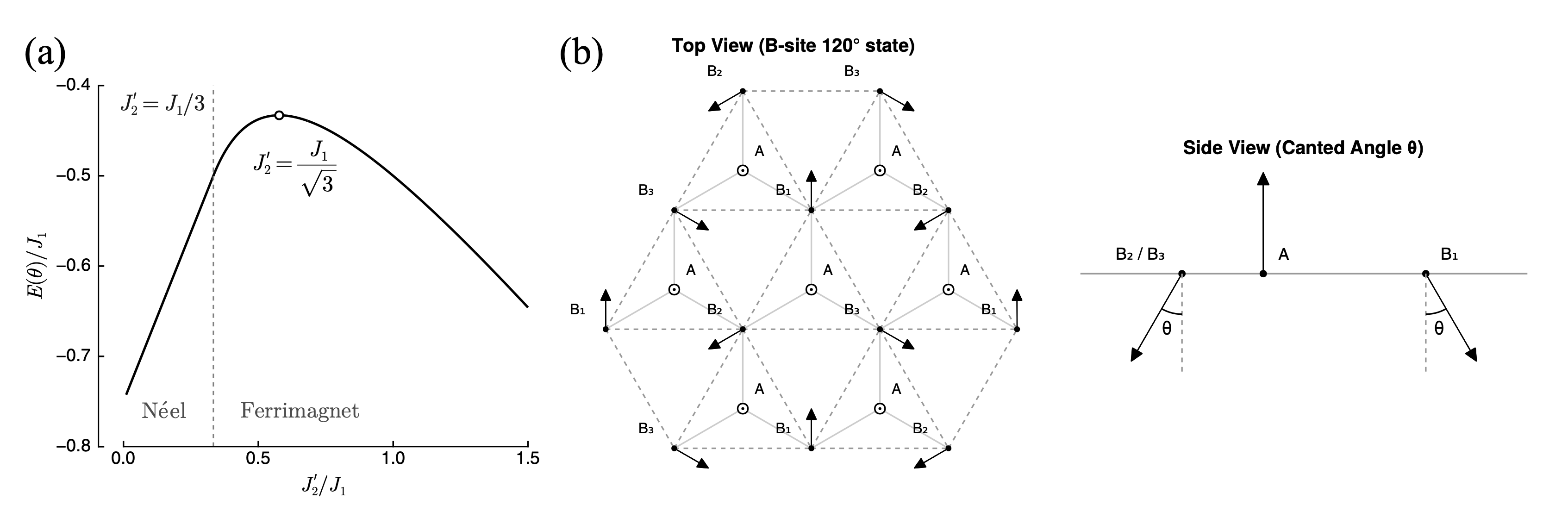}
    \caption{(a) Classical energy density per unit cell. The classical N\'eel-ferrimagnet transition is at $J_2' = J_1/3$. The flat-band point $J_2' = J_1/\sqrt{3}$ is also the maximum of the energy density. (b) A specific classical spin configuration with the lowest energy in the ferrimagnetic phase.}
    \label{fig:cl-spin}
\end{figure}

Here we give the details of the classical state discussed in the main text. For $J'_2\rightarrow 0$, the model reduces to the nearest-neighbor honeycomb Heisenberg antiferromagnet, whose classical ground state is the collinear N\'eel state and does not break translation symmetry. In the opposite limit $J_1\rightarrow 0$, the $B$ sites form a triangular-lattice antiferromagnet with $120^\circ$ order, while the $A$ spins are decoupled.

Between these limits, the two states can be continuously connected by the noncoplanar configuration shown in Fig.~\ref{fig:cl-spin}: the $B$ sublattice forms a canted $120^\circ$ pattern, and the $A$ spins are fully polarized. Parameterizing this family by the canting angle $\theta$, we obtain
\begin{align}
    \frac{4E(\theta)}{N_{uc}} = -3J_1\cos\theta + \frac{3J'_2}{2}\(3\cos^2\theta-1\).
\end{align}
Minimizing over $\theta$ gives a simple classical phase diagram. For $0< J'_2 < J_1/3$, the minimum is the honeycomb collinear N\'eel state. For $J'_2 > J_1/3$, the system prefers the canted noncollinear state with
\begin{align}
    \theta = \arccos\( \frac{J_1}{3J'_2} \).
\end{align}
We numerically checked that this canted state is always one member of the extensively degenerate classical ground-state manifold.
The flat-band point $J_1=\sqrt{3}J'_2$ also lies inside this noncollinear classical regime.

Coincidentally, we also find the variational $E(\theta)/N_{uc}$ has a maximum at the flat-band limit $J_1 = \sqrt{3}J'_2$. Another hint that this is an interesting model is that numerically, $E(J'_2 = 0.5)/N_{uc} = -0.4375J_1$, while the zero-point energy from linear spin-wave theory is $\Delta E/N_{uc}=-0.45826J_1$, showing anomalously large spin-wave corrections.

\section{Details of the iPEPS optimization and transfer-matrix calculations}

\begin{figure}[!h]
    \centering
    \includegraphics[width=0.92\linewidth]{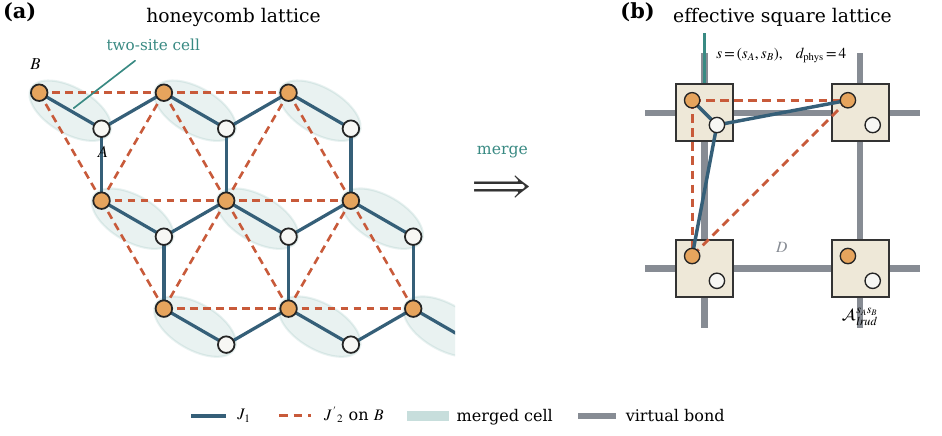}
    \caption{Mapping of the honeycomb model to the square-lattice iPEPS geometry. (a) The two spin-$1/2$ sites in each shaded primitive cell are grouped without truncating their local Hilbert space. Solid blue and dashed orange lines denote $J_1$ and the $B$-sublattice $J'_2$ couplings, respectively. (b) Each pair becomes a rank-five tensor $\mathcal{A}^{s_A s_B}_{lrud}$ with one composite physical index of dimension $d_{\mathrm{phys}}=2\times2=4$ and four virtual indices of dimension $D$. Gray lines form the effective square lattice. The colored bonds attached to a reference cell show the six Hamiltonian contributions used in one local-energy term: one intracell and two intercell $J_1$ bonds, and two axial plus one plaquette-diagonal $J'_2$ bonds.}
    \label{fig:ipeps-merge}
\end{figure}

To use a square-lattice iPEPS contraction, we combine the $A$ and $B$ spins of each primitive honeycomb cell into the composite physical index $s=(s_A,s_B)$, as shown in Fig.~\ref{fig:ipeps-merge}. This is an exact reshaping of the local Hilbert space rather than a physical-space truncation. The resulting tensor $\mathcal{A}^{s_A s_B}_{lrud}$ has four virtual legs of dimension $D$ and a composite physical dimension $d_{\mathrm{phys}}=4$. We use a real, one-tensor unit cell, which corresponds to the primitive two-site honeycomb unit cell. In the merged geometry, the local energy contains one intracell and two intercell $J_1$ interactions. The three $B$--$B$ interactions generated by $J'_2$ become two axial links and one diagonal across an effective square plaquette. Translating this local object tiles every bond of the original Hamiltonian once; the resulting energy is divided by the two physical spins in each merged cell to obtain the energy per site.

For a fixed iPEPS tensor, we form the double-layer row-to-row transfer operator and approximate its upper and lower dominant fixed points by separate uniform boundary MPSs of bond dimension $\chi$. Both boundaries are brought to mixed canonical form and solved with the general VUMPS algorithm~\cite{vumps}; combining the upper and lower fixed points then gives the environment used for reduced density matrices, energies, and observables. Depending on $D$ and $\chi$, each objective evaluation first performs up to $30$--$50$ non-differentiated VUMPS iterations, with target residuals between $10^{-10}$ and $10^{-8}$. The final four VUMPS updates remain in the differentiation graph. The internal fixed-point maps use five to eight power iterations during these differentiated updates, whereas final observables are reevaluated at fixed tensor with longer, $40$--$80$-step power solves.

Following the AD-VUMPS framework of Ref.~\cite{zhang2023adVUMPS}, we write the independent real tensor entries as a vector $x$ and evaluate $E(x)$ and its gradient by reverse-mode AD through the tensor contractions and the final VUMPS updates. Consequently, the gradient includes a finite-unrolled response of the boundary fixed points; it neither freezes the boundary throughout the derivative nor performs an implicit differentiation of the exactly converged fixed-point equations. The gradient is supplied to L-BFGS with a history of $200$ vectors and a Hager--Zhang line search allowing at most five energy/gradient evaluations per step. In the $D=4$--$8$ production sequence, we use a gradient tolerance of $10^{-7}$ and up to ten L-BFGS iterations at each $\chi$, together with a local norm-based normalization/restriction and the leading local-metric preconditioner introduced in Ref.~\cite{zhang2026precondition}. The environment dimension is increased in stages (typically $\Delta\chi=32$), reusing the converged boundary and optimizer history, and reaches $\chi=1024$ in the largest calculations. A calculation at the next $D$ is initialized by enlarging the preceding optimized tensor and adding a small random perturbation of amplitude $10^{-2}$. At the largest $D$ and $\chi$, checkpointed recomputation is used during the reverse pass to reduce memory without changing the objective. For the largest environments, the boundary-$\chi$ indices are block-distributed over a $2\times2$ MPI process grid and evaluated on four GPUs using the Slice2D contraction path; implementation and scaling details of this multi-GPU scheme will be reported elsewhere. All reported observables are evaluated only after fixing the optimized tensor and reconverging the boundary environment with the longer observable settings.

For the correlation-length scaling in Fig.~\ref{fig:peps-xi-tm}(a), we use the two leading eigenvalues of the converged PEPS transfer matrix,
\begin{align}
    \xi=-\frac{1}{\ln|\lambda_1/\lambda_0|}.
\end{align}
At each selected tensor bond dimension $D=4,5,6,7,8,10$, we retain the largest available converged transfer-matrix environment. The corresponding correlation lengths are $0.9370$, $1.1536$, $1.4955$, $1.8805$, $2.0175$, and $2.1775$, respectively. A free linear fit of $1/\xi$ against $1/D$ gives a slope $4.3653$, intercept $-0.0347\pm0.0643$, and $R^2=0.9713$. A fit constrained to $1/\xi=0$ at $1/D=0$ has slope $4.1714$, i.e. essentially the same line, reflecting an unconstrained intercept that is zero to within its uncertainty. The available data are therefore consistent with a diverging thermodynamic correlation length, but they do not establish one: the intercept depends on which bond dimensions enter the fit, moving from $-0.0896\pm0.0468$ for $D=4,5,6,8$ to $+0.1307\pm0.1059$ if only $D=6,7,8,10$ are used, because $1/\xi$ flattens at large $D$. The correlation lengths remain too short to exclude a crossover toward a small positive value.

\begin{figure}[t]
    \centering
    \includegraphics[width=0.62\linewidth]{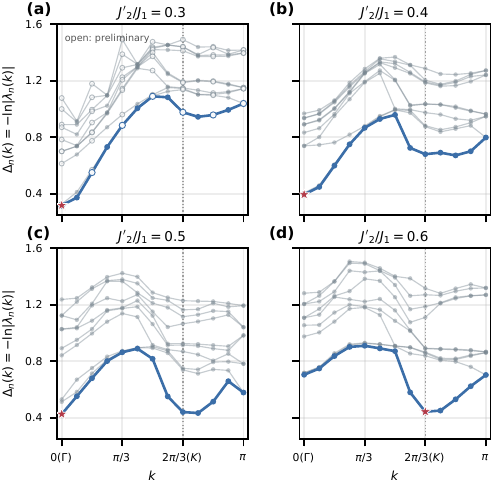}
    \caption{PEPS transfer-matrix spectra at $D=8$. The transfer-matrix gaps $\Delta_n(k)=-\ln|\lambda_n(k)|$ are shown for (a) $J'_2/J_1=0.3$, (b) $0.4$, (c) $0.5$, and (d) $0.6$. The lowest branch is highlighted in blue, gray curves show the higher branches, red stars mark the global minima, and the vertical dotted line marks $K$ at $k=2\pi/3$. Open markers in panel (a) denote preliminary older even-twelfth momentum points. The $K$ minimum progressively softens and becomes the global minimum at $J'_2/J_1=0.6$.}
    \label{fig:peps-tm-spectrum}
\end{figure}

For Fig.~\ref{fig:peps-mag}(c) and Supplemental Fig.~\ref{fig:peps-tm-spectrum}, we evaluate the one-sided transfer matrix of the $D=8$ merge-optimized PEPS with one fixed converged environment. The calculation uses dense tensors and does not project onto a selected spin-symmetry sector. The label \texttt{trivial} in the spectrum data denotes the excitation ansatz without a domain wall: the same boundary fixed point is used on both sides of the excitation. It does not denote restriction to the spin-singlet sector. The  momentum-resolved spectrum is calculated with the quasiparticle ansatz in the tangent space of the uniform boundary MPS~\cite{Haegeman2012Dispersion,Vanderstraeten2019Tangent}. We calculate the ten leading gaps $\Delta_n(k)=-\ln|\lambda_n(k)|$ at 13 momenta $k/\pi=0,1/12,\ldots,1$, using the same environment for all momenta at each coupling. The locations and values of the lowest gaps are summarized below. The $J'_2/J_1=0.3$ odd-twelfth points passed the stricter ten-level convergence gate, while its older even-twelfth points are shown as open markers. The $J'_2/J_1=0.4$ spectrum passed the same gate at all momenta. The $0.5$ and $0.6$ spectra use the original one-cycle Arnoldi semantics and return ten finite Ritz values at every momentum. This qualification is most relevant for the higher branches; the smooth lowest branch and its relocation from $\Gamma$ to $K$ are the qualitative features used in the main text, while quantitative level splittings require more converged Arnoldi calculations.

\begin{center}
    \small
    \centering
    \begin{tabular}{c c c}
    \hline
    $J'_2/J_1$ & momentum of minimum & $\Delta_{\mathrm{min}}$ \\
    \hline
    $0.3$ & $\Gamma$ & $0.3181$ \\
    $0.4$ & $\Gamma$ & $0.3947$ \\
    $0.5$ & $\Gamma$ & $0.4257$ \\
    $0.6$ & $K$ & $0.4428$ \\
    \hline
    \end{tabular}
\end{center}
Here $K$ corresponds to $k=2\pi/3$ in the momentum convention used in Figs.~\ref{fig:peps-mag}(c) and \ref{fig:peps-tm-spectrum}.

\section{Plaquette VBS variational comparison}
\label{sec:plaquette-vbs}

\begin{figure}[!t]
    \centering
    \includegraphics[width=0.96\linewidth]{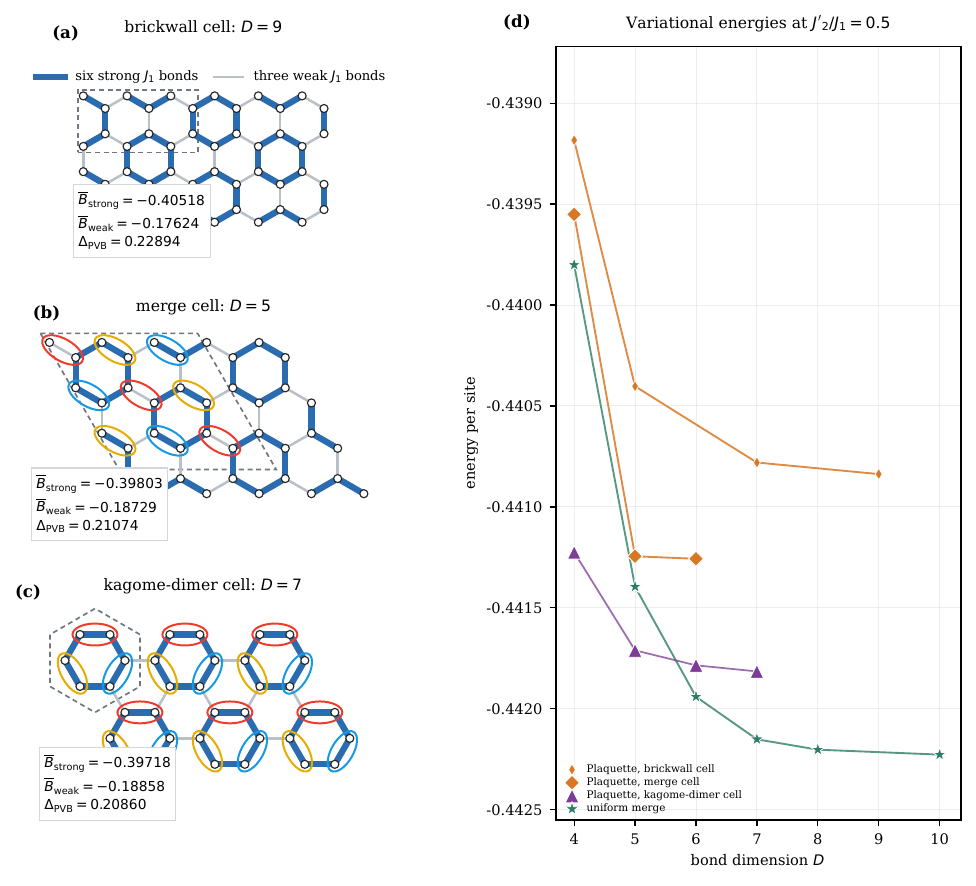}
    \caption{Plaquette-VBS (PVB) test at $J'_2/J_1=0.5$. Representative nearest-neighbor bond patterns for the PVB ansatz in (a) the brickwall cell, (b) the merge cell, and (c) the kagome-dimer cell. Thick blue and thin gray bonds indicate strong and weak $J_1$ correlations; colored ellipses mark merged spin pairs, and dashed boundaries indicate unit cells. (d) Energy comparison with the uniform ansatz. None of the three PVB constructions yields an energy below the best uniform result obtained in this study. Lines are guides to the eye.}
    \label{fig:plaquette-vbs-comparison}
\end{figure}

To test whether plaquette valence-bond (PVB) order is energetically favored at $J'_2/J_1=0.5$, we compare three PVB constructions with the translation-invariant uniform ansatz: the brickwall, merge, and kagome-dimer cells shown in Fig.~\ref{fig:plaquette-vbs-comparison}(a--c). The kagome network is obtained by merging pairs of honeycomb sites along a Kekul\'e dimer covering and is contracted using a one-hole embedding~\cite{Corboz2012Simplex}. The representative bond patterns retain a clear distinction between strong and weak bonds, showing that the PVB order survives optimization.

Among the results shown in Fig.~\ref{fig:plaquette-vbs-comparison}(d), the lowest energies per site for the brickwall, merge, and kagome-dimer PVB constructions are $-0.44083760$ ($D=9$), $-0.44125722$ ($D=6$), and $-0.44181506$ ($D=7$), respectively. All three remain above the best uniform result, $-0.44222719$ at $D=10$. The PVB energies show little change at the largest bond dimensions reached for each construction, indicating that the three energy sequences are nearly converged with increasing $D$. Thus, none of the PVB constructions improves upon the best uniform energy obtained in this study.

As a complementary test, we initialize a larger unit cell by tiling the optimized uniform tensor and reoptimize within the enlarged cell. This search likewise yields no PVB state with an energy below the uniform result. Together with the nearly saturated PVB energy sequences, this supports the absence of an energetically favored PVB state at $J'_2/J_1=0.5$.

For the kagome-dimer and uniform branches, the comparison uses results obtained at the largest available environment dimensions.

\section{Details on symmetric MPS calculations on the cylinder}

Here we give more details on the cylinder MPS calculations. We keep the full SU(2) symmetry for these calculations, except for the flux insertion part, where only the U(1) subgroup is kept. The bond dimension $D$ in this section will be the effective dense bond dimension. We use the variational uniform matrix product state (VUMPS) algorithm \cite{vumps, devos_2026_20144080, haegeman2024tensorkit} and conjugate gradient when VUMPS convergence is slow \cite{conjugateGradient}. We always keep the convergence tolerance to be below $10^{-6}$.

\begin{figure}
    \centering
    \includegraphics[width=\linewidth]{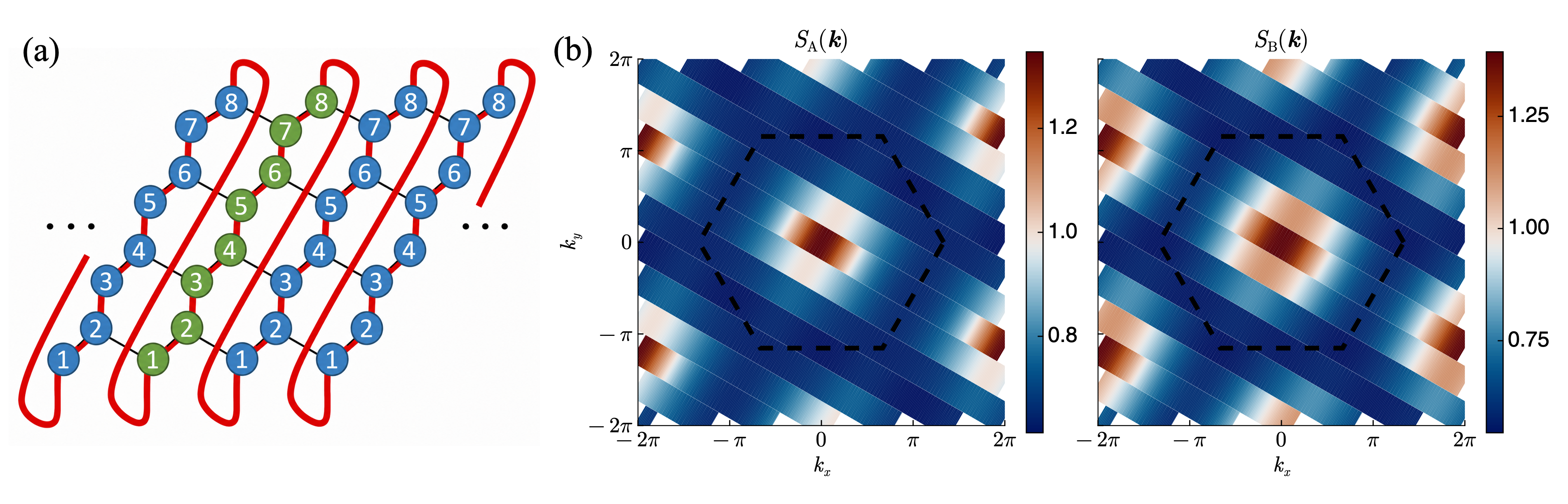}
    \caption{(a) YC8 cylinder geometry and one-dimensional MPS ordering. The green sites indicate one MPS unit cell. (b) Sublattice-resolved spin structure factors on YC12 at $J_2'=0.5J_1$ and $D=6400$. The only visible maximum occurs at $\Gamma$, with no finite-momentum peak indicating magnetic translation-symmetry breaking.}
    \label{fig:mps-ssf}
\end{figure}

We use the YC cylinder geometry for the MPS calculations since this allows for a single-column unit cell. Fig.~\ref{fig:mps-ssf}(a) shows the one-dimensional MPS ordering on a representative YC8 cylinder. In practice, we have performed calculations on YC6, 8, 10, and 12 cylinders, with the largest bond dimension capped at $D=9600$. We will always use $J_2'=0.5J_1$ as the example state for the symmetric phase described in the main text.

Fig.~\ref{fig:mps-ssf}(b) shows that the sublattice-resolved spin structure factors have no finite-momentum peak, suggesting the absence of magnetic translation-symmetry breaking. We further check order parameters that are not directly visible in the spin structure factor. Fig.~\ref{fig:mps-corr} shows representative unconnected correlations of the nearest-neighbor bond energy and scalar chirality. Both correlations decay exponentially with short fitted correlation lengths, giving no evidence for long-range valence-bond or chiral order on the cylinder. We also explicitly check that $\<\mathbf{S}_1\cdot(\mathbf{S}_2\times\mathbf{S}_3)\>$ vanishes for all the optimized MPSs; hence, we find no evidence for time-reversal symmetry breaking.

\begin{figure}
    \centering
    \includegraphics[width=\linewidth]{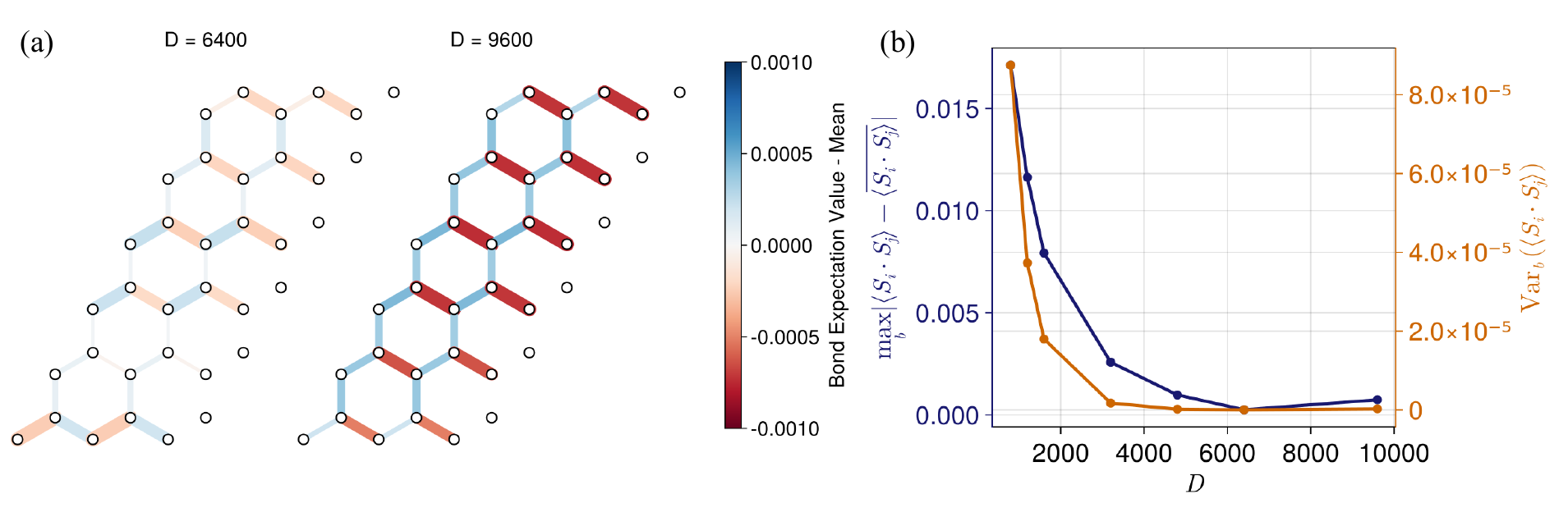}
    \caption{Bond expectation values in the optimized YC12 MPS at $J_2'=0.5J_1$. (a) Deviation $\langle \mathbf{S}_i\cdot\mathbf{S}_j\rangle-\overline{\langle \mathbf{S}_i\cdot\mathbf{S}_j\rangle}$ from the mean over nearest-neighbor bonds in the MPS unit cell, shown for $D=6400$ and $D=9600$. (b) Maximum deviation and variance versus $D$. The variance is rapidly suppressed, while the residual maximum deviation remains small on the scale of the bond energy.}
    \label{fig:mps-bonds}
\end{figure}

As a complementary local diagnostic for lattice symmetry breaking, we compare the bond expectation values in the optimized MPS. Fig.~\ref{fig:mps-bonds} shows that the variance of the bond deviations is rapidly suppressed to below $10^{-4}$ with increasing bond dimension for YC12. The largest residual deviation is not strictly monotonic in $D$, but it remains small on the scale of the bond energy and does not develop into a stable enlarged-unit-cell pattern. Although the cylinder geometry explicitly distinguishes the two lattice directions, this anisotropy is strongly suppressed in the large-$D$ YC12 bond pattern, supporting the interpretation that the wide-cylinder calculations capture some 2D physics.

\begin{figure}
    \centering
    \includegraphics[width=\linewidth]{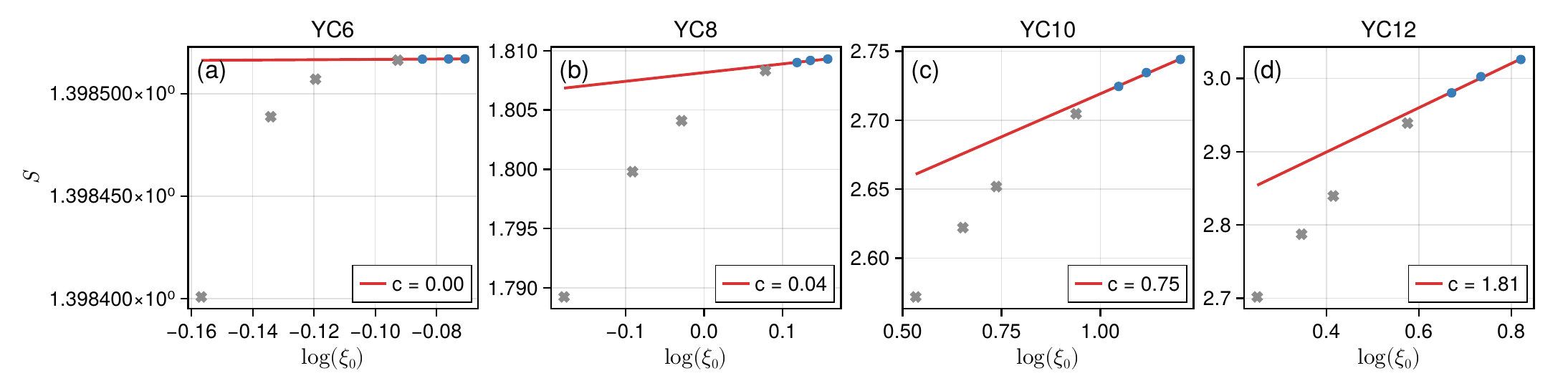}
    \caption{Finite-entanglement scaling of the cylinder MPS in the $S=0$ sector, using bond dimensions $D=[800,1200,1600,3200,4800,6400,9600]$. For YC10 and YC12 the entropy continues to grow over the accessible range of $\xi_0$, showing that these wider cylinders are not fully entanglement-saturated at the available bond dimensions. The red lines are linear fits to the largest-$D$ points and give effective central charges for the finite-width cylinders.}
    \label{fig:mps-ee}
\end{figure}

We also examine the finite-entanglement behavior of the cylinder MPS. As shown in Fig.~\ref{fig:mps-ee}, the narrower YC6 and YC8 cylinders are close to saturation over the accessible bond dimensions, while YC10 and YC12 continue to show visible growth of the entropy with the transfer-matrix correlation length. Nevertheless, the fitted slopes decrease with increasing bond dimension. In particular, the effective central charge extracted from the three largest-$D$ points is below one for YC10, which is consistent with eventual saturation on the finite-width cylinder rather than a stable critical state, which could be due to the emergent gauge field dynamically changing the boundary conditions of spinons \cite{He2017}.

\begin{figure}
    \centering
    \includegraphics[width=\linewidth]{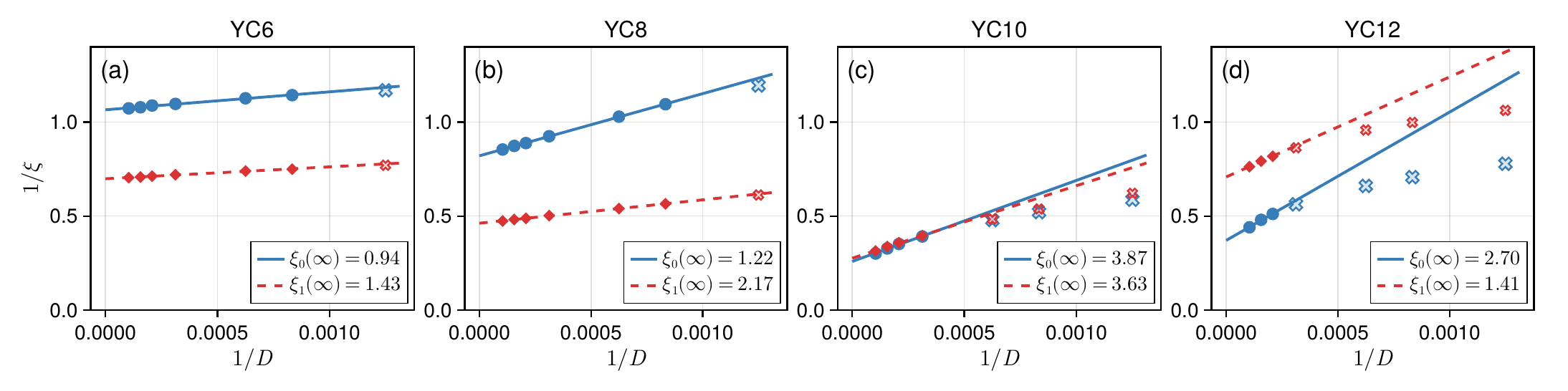}
    \caption{Inverse correlation lengths in the $S=0$ ($\xi_0$) and $S=1$ ($\xi_1$) sectors as functions of $1/D$ for YC6, YC8, YC10, and YC12 cylinders. Filled markers denote the large-$D$ data points used in the linear $1/D$ fits, while open cross markers show smaller-$D$ points not included in the fits. The red solid and orange dashed lines are the fits for the $S=0$ and $S=1$ sectors, respectively; the legends report the extrapolated correlation lengths in the $D\rightarrow\infty$ limit.}
    \label{fig:mps-xi}
\end{figure}

With this point in mind, we also examine the leading correlation lengths from transfer matrices of the optimized MPSs, shown in Fig.~\ref{fig:mps-xi}. Similar to the entanglement entropy scaling, the narrower YC6 and YC8 cylinders show good fits and convergence in both the $S=0$ ($\xi_0$) and $S=1$ ($\xi_1$) sectors, while only the large-$D$ points in YC10 and YC12 are more linear. Using the $1/D$ extrapolations, we find all the correlation lengths to be finite in the $D\rightarrow\infty$ limit. The other commonly used extrapolation scheme conjectured in \cite{marekgap} produces poorer fits, likely because the wider-cylinder MPSs are not fully converged in bond dimension. Fig.~\ref{fig:mps-xi} also shows an unusual feature: the lowest-energy excitations change sector from $S=1$ to $S=0$. In particular, YC12 cylinders have $\xi_0>\xi_1$ for all values of $D$ we computed. YC10 cylinders sit close to this excitation-sector crossover and have larger correlation lengths than other widths. Since the momentum cuts associated with the cylinder geometry are not special, we interpret the larger correlation lengths as being more closely related to this 1D--2D crossover than to the true 2D physics.

\begin{figure}
    \centering
    \includegraphics[width=0.8\linewidth]{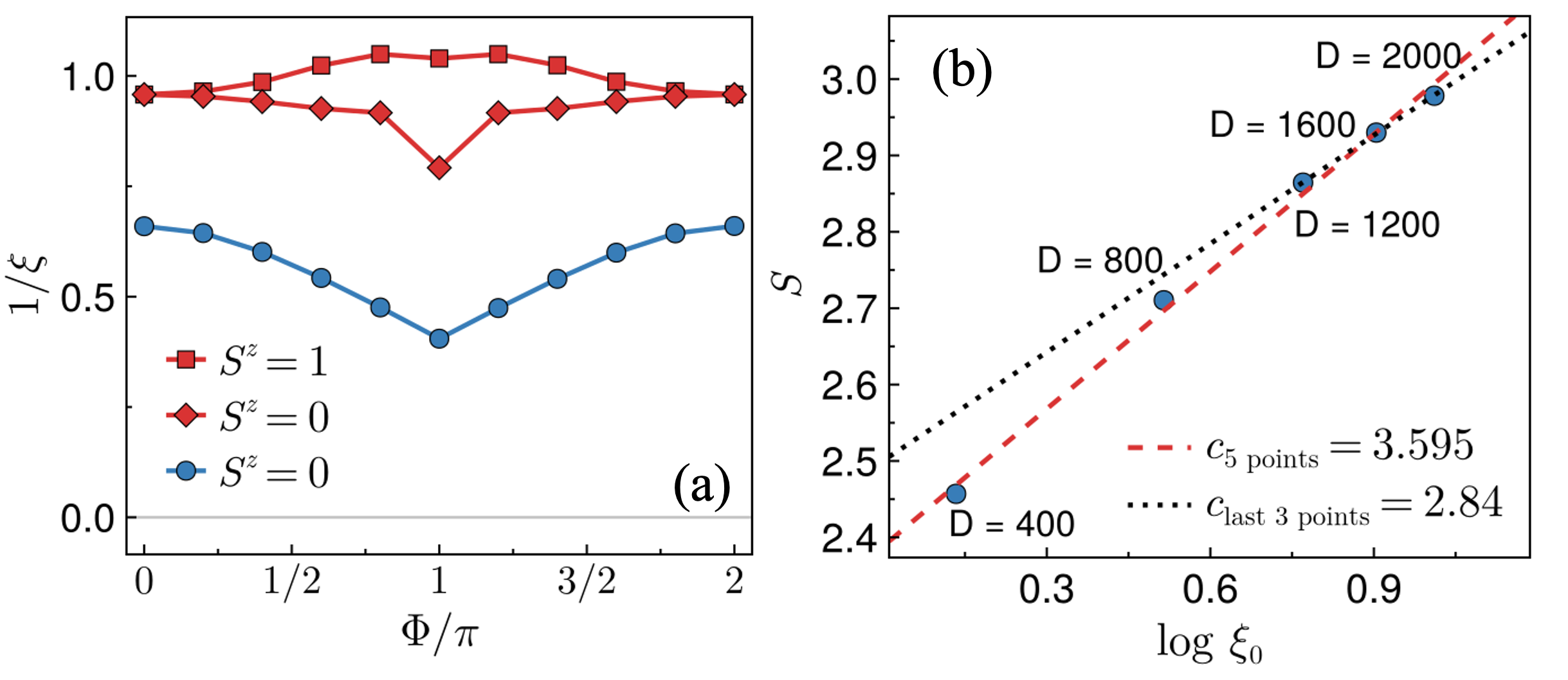}
    \caption{Flux-threading response at $J_2'/J_1=0.5$ on a YC12 cylinder, obtained with U(1)-symmetric MPS. (a) Inverse correlation lengths associated with the leading and subleading $S^z=0$ transfer-matrix eigenvalues and the leading $S^z=1$ eigenvalue at $D=1600$. At zero flux, the subleading $S^z=0$ mode (red diamonds) is degenerate with the leading $S^z=1$ mode (red squares), identifying them as components of the leading $S=1$ triplet. The unmatched leading $S^z=0$ mode (blue circles) therefore belongs to the $S=0$ sector. Flux reduces SU(2) to U(1) and splits the triplet components, while both tracked $S^z=0$ correlation lengths are maximized at $\Phi=\pi$. On the narrower cylinders, the leading $S^z=0$ mode instead belongs to the $S=1$ triplet, while the total spin of the next $S^z=0$ mode cannot be determined by this degeneracy criterion. (b) Entanglement entropy $S$ versus $\log\xi_0$ at $\Phi=\pi$ for $D=400$--$2000$. Fitting all five points gives $c_{\mathrm{eff}}\approx3.6$, whereas fitting the largest three gives $c_{\mathrm{eff}}\approx2.8$, revealing a downward drift with bond dimension.}
    \label{fig:YC12flux}
\end{figure}

For the YC12 cylinders, similar to the Dirac spin liquid calculations \cite{PhysRevX.7.031020, PhysRevLett.123.207203}, the flux threading enhances the leading $S^z=0$ correlation lengths, with the $\pi$-flux state showing substantial finite-entanglement growth, as shown in Fig.~\ref{fig:YC12flux}(a), which can be interpreted as adiabatically approaching the Dirac points via flux pumping. The entanglement entropy versus correlation length at $\pi$-flux is shown in Fig.~\ref{fig:YC12flux}(b), where we fit to the standard finite-entanglement scaling $S = c/6 \log \xi_0 +$ const. and find an effective central charge $c \approx 3.6$. This is consistent with a gapless state, but the effective central charge is smaller than expected for the free Dirac spinon scenario \cite{zhu2018entanglement, song2019unifying, PhysRevX.10.011033}.

To further examine whether the symmetric state is a symmetry-protected topological (SPT) phase, we also compute the strange correlator \cite{PhysRevLett.112.247202},
\begin{align}
    C(r) = \frac{\<\Omega|\phi(0)\phi(r)|\Psi\>}{\<\Omega|\Psi\>}.
    \label{eq:corr}
\end{align}
Here $|\Psi\>$ is a potential SPT state being tested, while $|\Omega\>$ is a trivial product state respecting all spatial symmetries of the problem. Normally $\<\Omega|\phi(0)\phi(r)|\Psi\>$ decays to zero exponentially fast since the two states are orthogonal, but since the denominator in Eq.~\eqref{eq:corr} decays in the same fashion, this quantity is useful in discerning whether a state is a strong SPT state \cite{PhysRevB.93.245141, gao2026spurious}.

\begin{figure}
    \centering
    \includegraphics[width=0.5\linewidth]{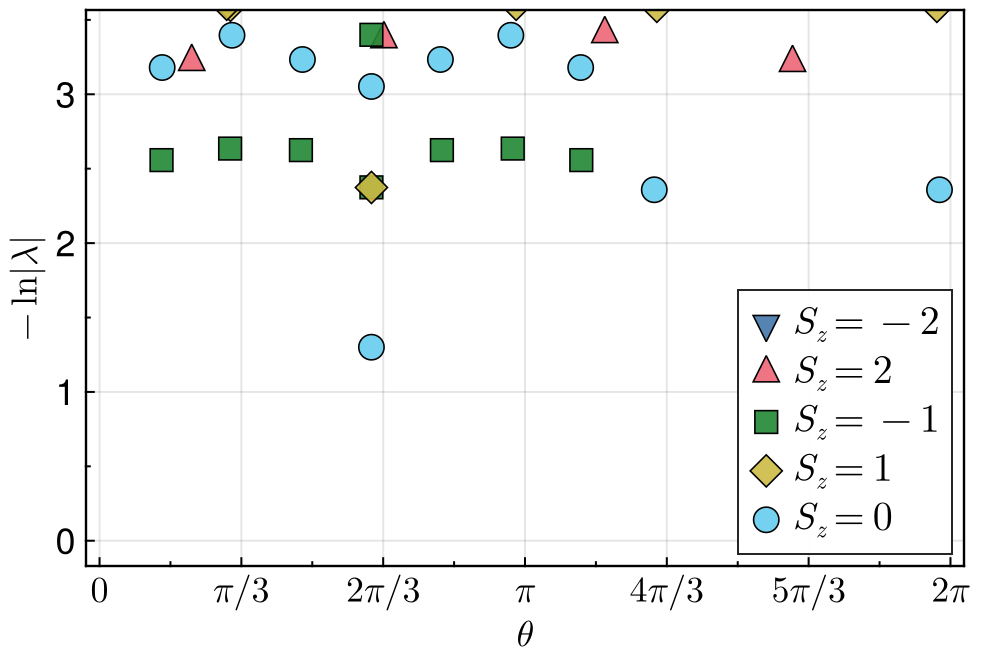}
    \caption{Mixed transfer matrix spectrum between a staggered product state and the ground state of our model at $J_2'=0.5J_1$ on a YC12 cylinder with $D=1600$.}
    \label{fig:mps-strangeCo}
\end{figure}

In our case, $|\Psi\>$ is the optimized YC12 symmetric state at $J_2'=0.5J_1$ and $D=1600$. To avoid the Heisenberg model on an even-width cylinder producing an SPT phase, we lower the symmetry to U(1) and produce the product state $|\Omega\>$ using only a staggered-field Hamiltonian. We explicitly check that $|\Omega\>$ is completely trivial in its entanglement spectrum content. To obtain Eq.~\eqref{eq:corr}, we only need to compute the spectrum of the mixed transfer matrix $\mathbb{E} = \sum_i \bar{A}^i_{\Omega}\otimes A^i_{\Psi}$, shown in Fig.~\ref{fig:mps-strangeCo}. We represent the spectrum as $\lambda = |\lambda| e^{i\theta}$. Note that the leading eigenvalue $|\lambda_0| < 1$, consistent with the two states being orthogonal. The decay behavior of $C(r)$ is then determined by $|\lambda_1|/|\lambda_0| < 1$. In other words, the strange correlator in our case is exponentially decaying, ruling out a strong SPT phase.

\begin{figure}
    \centering
    \includegraphics[width=\linewidth]{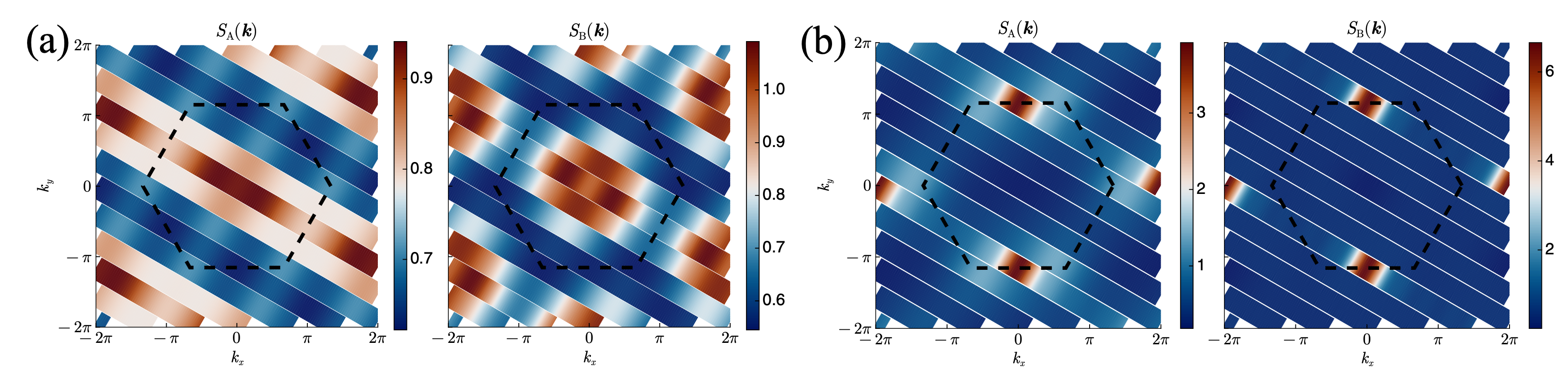}
    \caption{Sublattice-resolved spin structure factors $S_A(\mathbf{k})$ and $S_B(\mathbf{k})$ on a YC12 cylinder with $D=1600$ at (a) $J'_2/J_1=0.6$ and (b) $J'_2/J_1=2$. Only the momentum cuts accessible on YC12 are shown; dashed hexagons mark the first Brillouin zone. At $J'_2/J_1=0.6$, the dominant correlations begin to shift away from $\Gamma$, with distinct momentum-space distributions on the two sublattices. At $J'_2/J_1=2$, both sublattice structure factors peak at the $M$ points, consistent with stripe antiferromagnetic correlations.}
    \label{fig:largeJ2p}
\end{figure}

Finally, we examine the evolution of the spin correlations beyond the symmetric regime. At $J'_2/J_1=0.6$, Fig.~\ref{fig:largeJ2p}(a) shows that the dominant response begins to shift away from $\Gamma$, with distinct momentum-space distributions on the $A$ and $B$ sublattices. This sublattice-dependent evolution is consistent with the onset of the finite-wave-vector regime inferred from the ED calculations in the following section. By $J'_2/J_1=2$, both sublattice structure factors peak at the $M$ points [Fig.~\ref{fig:largeJ2p}(b)], indicating stripe antiferromagnetic correlations. This differs qualitatively from the classically expected canted ferrimagnetic state [Fig.~\ref{fig:cl-spin}(b)]. The preference for collinear correlations suggests that quantum fluctuations remain important even in this large-$J'_2$ regime.

\section{Details on exact diagonalization}

In this section we give more details on exact diagonalization (ED) calculations on $N=24$-site clusters, which are compatible with the symmetries of the expected valence-bond crystals from the usual Heisenberg models on the honeycomb lattice \cite{PhysRevB.84.024406}. A larger cluster that is compatible with these symmetries is $N=42$, making the computation of observables substantially more expensive; this is left for future work. To complement the PEPS phase diagram with $J_2' < 0.5 J_1$, we use tower-of-states diagnostics to check for phase transitions out of the N\'eel state. We then examine the spin structure factors for $J_2' \geq 0.5 J_1$, where ED provides additional evidence for competing phases beyond the symmetric regime discussed above.

\subsection{Tower of states for $J_2' \leq 0.5 J_1$}

\begin{figure}
    \centering
    \includegraphics[width=\linewidth]{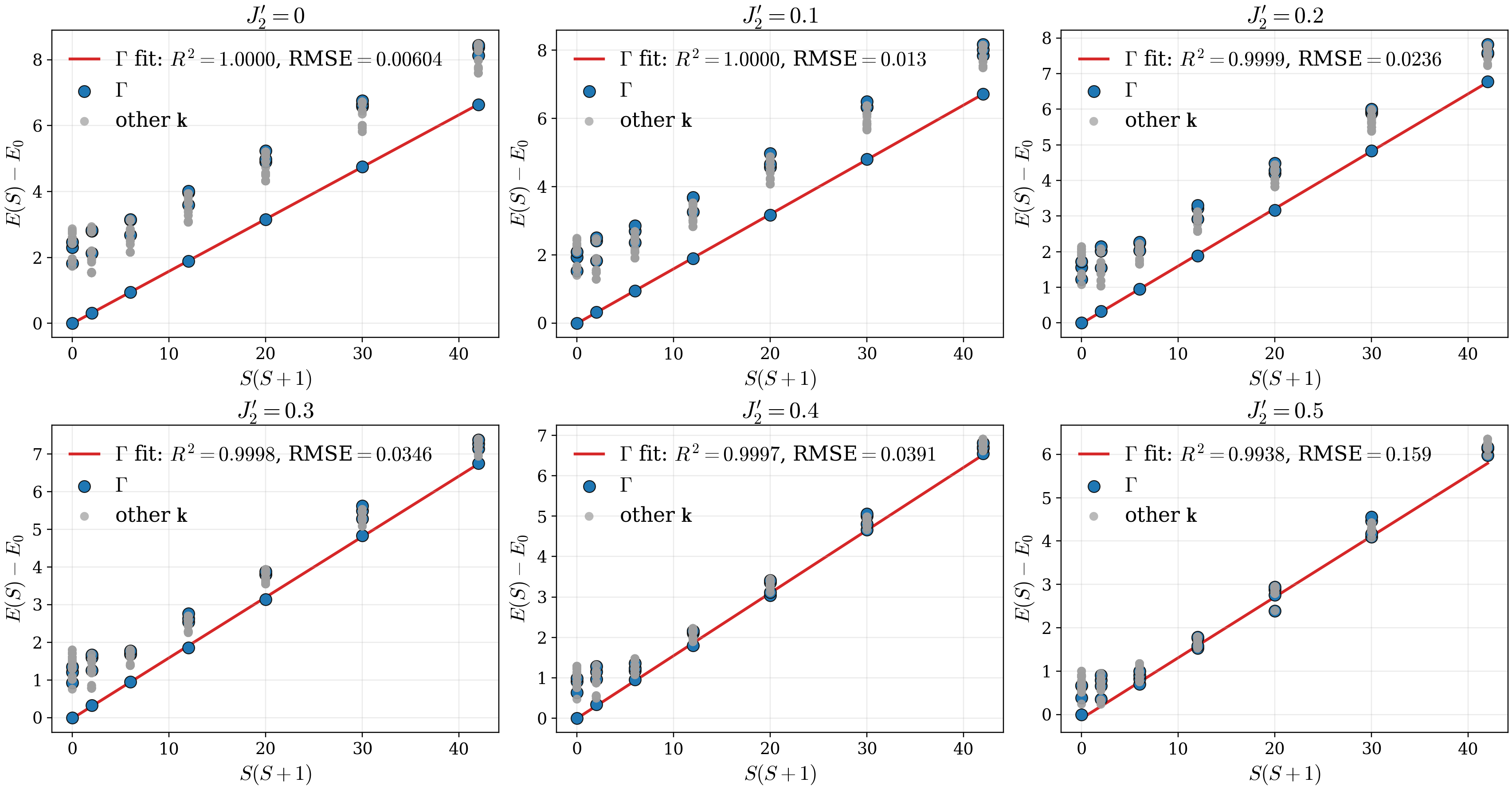}
    \caption{Low-energy spectra from $N=24$ cluster ED calculations, plotted as a function of $S(S+1)$ for $J_2'/J_1=0,0.1,\ldots,0.5$. Blue points denote the lowest state in each total-spin sector at $\Gamma$, while gray points denote low-lying states at other momenta. The red line is a linear fit to the lowest-energy $\Gamma$ states, as expected for an Anderson tower of states in a N\'eel ordered phase. Clear deviations from this linear tower behavior appear starting near $J_2'/J_1=0.4$.}
    \label{fig:ED-tos}
\end{figure}

Here we perform a simple tower-of-states analysis to distinguish a N\'eel ordered state at small $J_2'$ from the symmetric state at larger $J_2'$, without the full point group analysis. For a finite cluster whose thermodynamic ground state has N\'eel order, the low-energy spectrum contains an Anderson tower of states in different total-spin sectors, with energies that approximately follow the quantum-rotor form $E_S-E_0\propto S(S+1)/N$. We therefore inspect whether the low-energy spectrum on the $N=24$ cluster shows this N\'eel tower structure as $J_2'$ is increased.

Fig.~\ref{fig:ED-tos} shows the low-energy spectrum organized by total spin for $J_2' \in [0,0.5]$. At small $J_2'$, the lowest $\Gamma$ state in each spin sector follows the expected linear dependence on $S(S+1)$, while states at other momenta remain higher in energy than this tower. This behavior is consistent with the N\'eel ordered regime in the usual $J_1$ model on the honeycomb lattice. As $J_2'$ increases, the linear tower structure deteriorates and the separation from other momentum sectors is reduced, indicating the loss of N\'eel order. The tower structure is most clearly lost by $J_2'/J_1=0.5$, consistent with the symmetric regime found in the PEPS and MPS calculations.

\begin{figure}
    \centering
    \includegraphics[width=\linewidth]{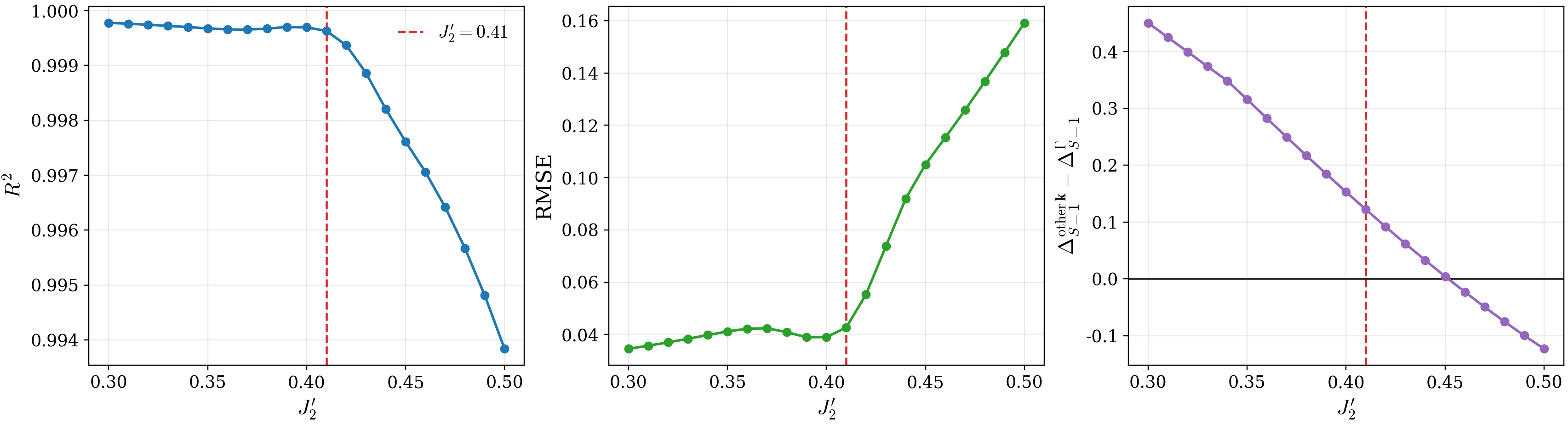}
    \caption{Fine scan of the tower-of-states diagnostics between $J_2'/J_1=0.3$ and $0.5$ with step size $0.01$. The left and middle panels show the $R^2$ and root-mean-square error (RMSE) of the linear fit to the $\Gamma$ tower states in Fig.~\ref{fig:ED-tos}. While $R^2$ remains close to one, the RMSE resolves the absolute size of the fit residuals and shows the clearest change of behavior near the dashed red line at $J_2'/J_1=0.41$. The right panel shows the energy difference in the $S=1$ sector between the lowest state at nonzero momentum and the lowest state at $\Gamma$; its sign change near $J_2'/J_1=0.45$ indicates that the lowest $S=1$ excitation changes momentum.}
    \label{fig:ED-tos-fit}
\end{figure}

To locate this boundary more precisely, Fig.~\ref{fig:ED-tos-fit} tracks the quality of the linear tower fit in a finer scan between $J_2'/J_1=0.3$ and $0.5$. Since $R^2$ remains close to one throughout this range, we also plot the root-mean-square error (RMSE), which measures the absolute residual of the tower fit in energy units. The fit quality changes most clearly near $J_2'/J_1\simeq 0.41$, where the RMSE begins to increase rapidly. A sharper finite-size diagnostic comes from the $S=1$ sector: near $J_2'/J_1\simeq 0.45$, the lowest triplet excitation switches from the $\Gamma$ sector to a nonzero momentum sector. This switching clearly marks the loss of the usual N\'eel order, which orders at $\Gamma$ for the honeycomb lattice. From this evidence, we estimate the N\'eel-to-symmetric transition to occur near $J_2'/J_1\simeq 0.4$ on the $N=24$ cluster, which is numerically close to the PEPS estimate.

\subsection{Spin structure factors for $J_2' \geq 0.5 J_1$}

\begin{figure}
    \centering
    \includegraphics[width=\linewidth]{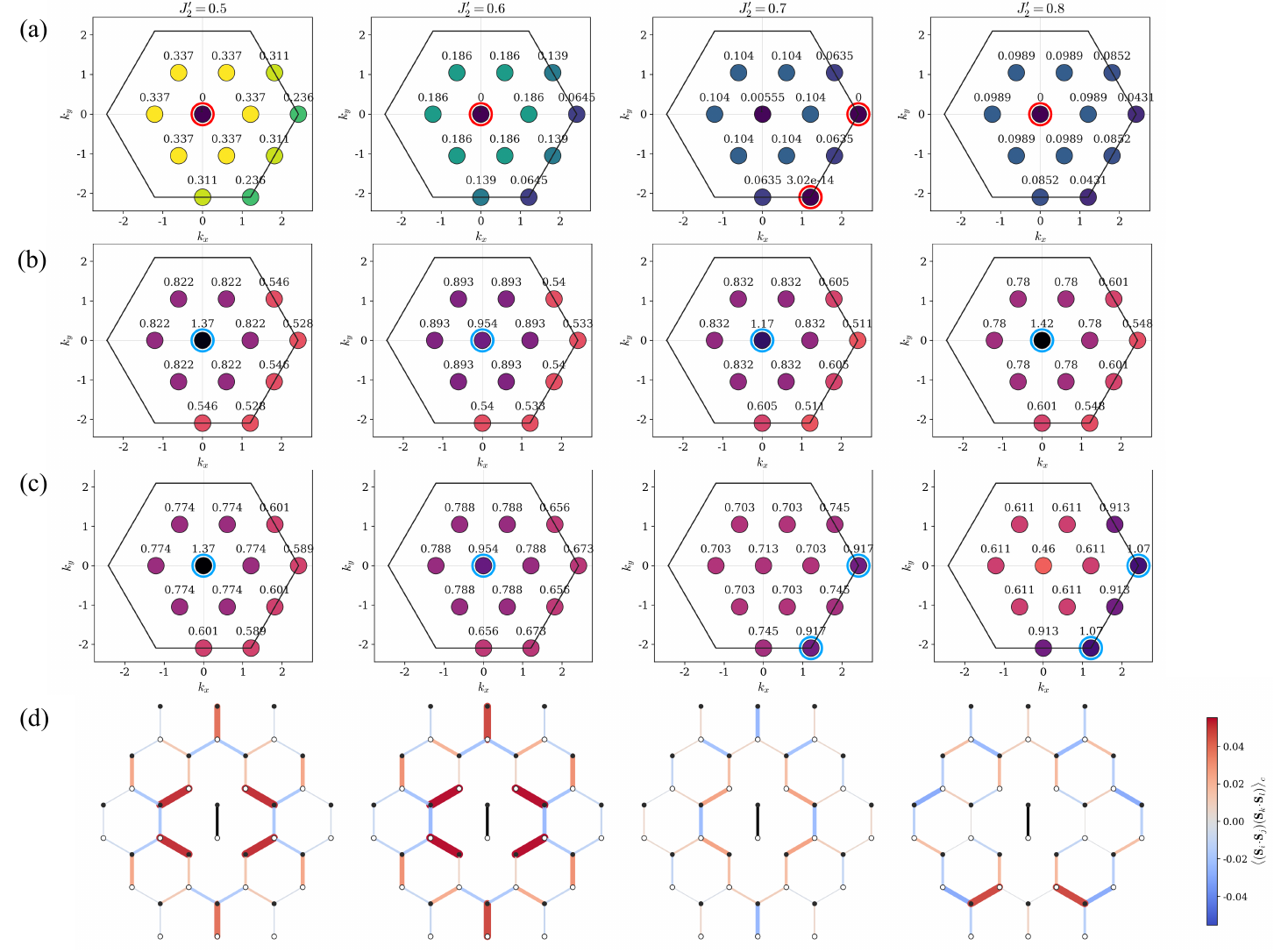}
    \caption{ED diagnostics on the $N=24$ cluster for $J_2'/J_1=0.5,0.6,0.7,0.8$. (a) Lowest energy in each momentum sector relative to the global ground-state energy. Red circles mark the lowest-energy momentum sector(s). The ground state is at $\Gamma$ for $J_2'/J_1=0.5,0.6,0.8$, while nearly degenerate finite-momentum states appear near $J_2'/J_1=0.7$. (b,c) Sublattice-resolved spin structure factors on the $A$ and $B$ sublattices, respectively. Blue circles mark the dominant structure-factor peak(s); the peak remains at $\Gamma$ on the $A$ sublattice but shifts to finite momentum on the $B$ sublattice for $J_2'\gtrsim 0.7J_1$. (d) Connected bond-bond correlations in real space, with the black bond indicating the reference bond. The $J_2'/J_1=0.5,0.6$ states show weak, rapidly decaying plaquette-like correlations, while the $J_2'/J_1=0.7,0.8$ correlations are weaker and less commensurate with the $N=24$ cluster.}
    \label{fig:ED-SSF}
\end{figure}

We consider $0.5\leq J_2'/J_1\leq 1$, where PEPS and symmetric MPS calculations indicate the onset of finite-momentum spin correlations. We corroborate this with ED calculations on the $N=24$ cluster. This calculation puts the upper boundary of the symmetric phase (zero-momentum regime) at $J_2'/J_1=0.61$.

Fig.~\ref{fig:ED-SSF} summarizes the ED observations at representative values $J_2'/J_1=0.5,0.6,0.7,0.8$. The lowest state remains in the $\Gamma$ momentum sector at $J_2'/J_1=0.5$ and $0.6$, consistent with the symmetric regime discussed above. Around $J_2'/J_1=0.7$, low-lying finite-momentum states become competitive, with only a small energy advantage to the $\Gamma$ sector. The sublattice-resolved structure factors show that the finite-momentum response is mainly associated with the $B$ sublattice in this regime, consistent with the finite-momentum pattern expected from the classical state in Fig.~\ref{fig:cl-spin}. Around $J_2'/J_1=0.8$, the lowest state returns to the $\Gamma$ momentum sector while retaining a similar finite-momentum structure factor. The real-space bond correlations remain weak throughout this range, without a robust valence-bond pattern on the $N=24$ cluster.

\begin{figure}
    \centering
    \includegraphics[width=0.8\linewidth]{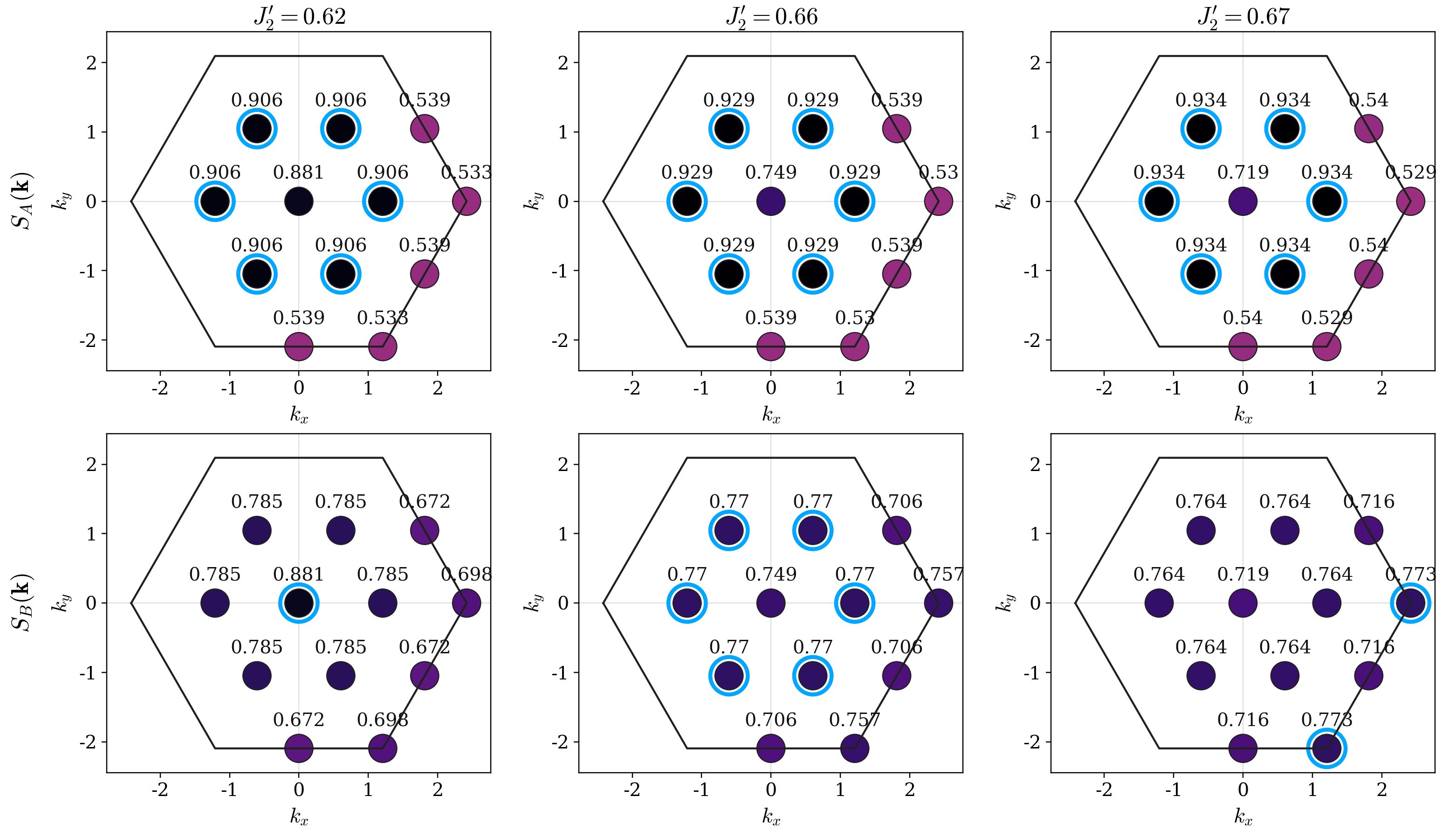}
    \caption{Representative sublattice-resolved spin structure factors on the $N=24$ cluster for $J_2'/J_1=0.62,0.66,0.67$. The ground state remains in the $\Gamma$ momentum sector for these points. Blue circles mark the dominant peak positions. The $A$-sublattice structure factor has dominant peaks at finite momenta in this range, while the $B$-sublattice peak evolves from $\Gamma$ at $J_2'/J_1=0.62$ to finite momenta as $J_2'$ approaches the regime shown in Fig.~\ref{fig:ED-SSF}.}
    \label{fig:ED-ssf-fine}
\end{figure}

To resolve how the finite-momentum features emerge between the coarse points in Fig.~\ref{fig:ED-SSF}, we perform a finer scan between $J_2'/J_1=0.6$ and $0.7$. Fig.~\ref{fig:ED-ssf-fine} shows representative points at $J_2'/J_1=0.62,0.66,0.67$, where the ground state still lies in the $\Gamma$ sector. The $A$-sublattice structure factor already has dominant finite-momentum peaks in this interval, while the dominant $B$-sublattice peak shifts away from $\Gamma$ only closer to $J_2'/J_1\simeq 0.67$. This suggests that the finite-momentum regime develops separately on the two sublattices, making the physics in this range difficult to understand. The finite-momentum mesh is particularly coarse on system sizes available to ED. However, the different finite-momentum peaks on the two sublattices also make it difficult to perform infinite PEPS calculations in this regime.

\subsection{Phase diagram from ED}

\begin{figure}[!h]
    \centering
    \includegraphics[width=\linewidth]{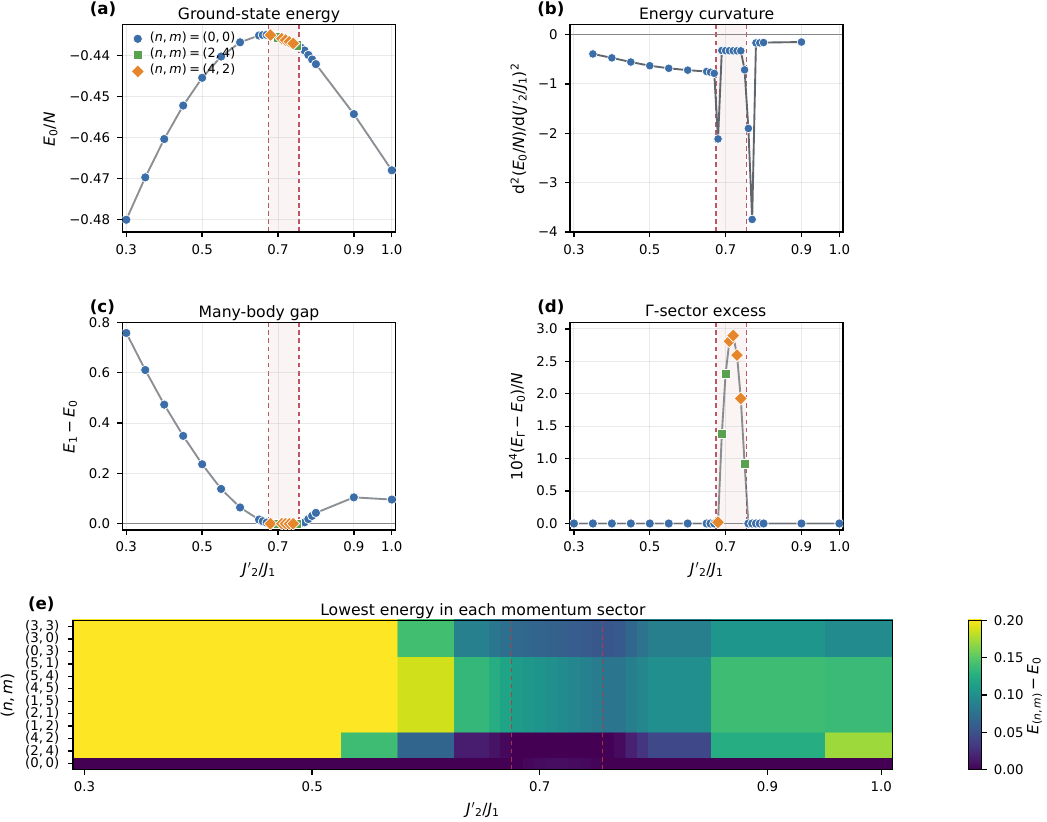}
    \caption{Cluster-resolved ED phase diagram for the $N=24$ honeycomb torus,
    obtained by scanning all 12 momentum sectors.  (a) Ground-state energy per
    site, with colors and marker shapes denoting the ground-state momentum.
    The dashed lines are the midpoints $J'_2/J_1=0.675$ and $0.755$ between
    adjacent scan points that bracket the two level crossings; the lightly
    shaded region is the finite-momentum ground-state window.  (b) Numerical
    second derivative of $E_0/N$ on the nonuniform coupling grid.  (c) Gap
    $E_1-E_0$ across all momentum sectors.  Its near-zero value in the shaded
    window reflects the nearly degenerate time-reversal pair $(2,4)$ and
    $(4,2)$ on this cluster.  (d) Energy density of the lowest $\Gamma$ state
    relative to the true ground state.  (e) Lowest energy in every momentum
    sector relative to $E_0$; the color scale is capped at $0.2J_1$.  Momentum
    labels use the same reciprocal-lattice indices $(n,m)$ as
    Fig.~\ref{fig:ED-ssf-fine}, so the same pair denotes the same physical
    $\mathbf{k}$ point in both figures; lines are guides to the eye.}
    \label{fig:ED-phase-diagram}
\end{figure}

We finally summarize the cluster-resolved evolution over the full interval
$0.3\leq J'_2/J_1\leq1$ by diagonalizing all 12 momentum sectors of the
$N=24$ torus.  As shown in Fig.~\ref{fig:ED-phase-diagram}, the ground state
remains in the $\Gamma$ sector through $J'_2/J_1=0.67$, moves to the
time-reversal-related sectors $(n,m)=(2,4)$ and $(4,2)$ for
$0.68\leq J'_2/J_1\leq0.75$, and returns to $\Gamma$ at
$J'_2/J_1=0.76$.  The finite-cluster level crossings are therefore bracketed
by $0.67<J'_2/J_1<0.68$ and $0.75<J'_2/J_1<0.76$.  The sharp features in
the energy curvature and the momentum-sector heat map independently expose
the same rearrangements.

The alternation between $(2,4)$ and $(4,2)$ inside the shaded window is the
choice between a time-reversal pair and does not represent additional phase
boundaries.  Likewise, the near-vanishing finite-cluster gap in
Fig.~\ref{fig:ED-phase-diagram}(c) should not be interpreted as direct
evidence for a vanishing bulk gap.  These crossing locations are specific to
the $N=24$ cluster: the structure-factor evolution discussed above begins
already near $J'_2/J_1\simeq0.6$, before the finite-momentum sector becomes
the absolute ground state.

\section{VUMPS contraction of trial state PEPS on honeycomb lattice}
\begin{figure}[!h]
    \centering
    \includegraphics[width=0.39\linewidth]{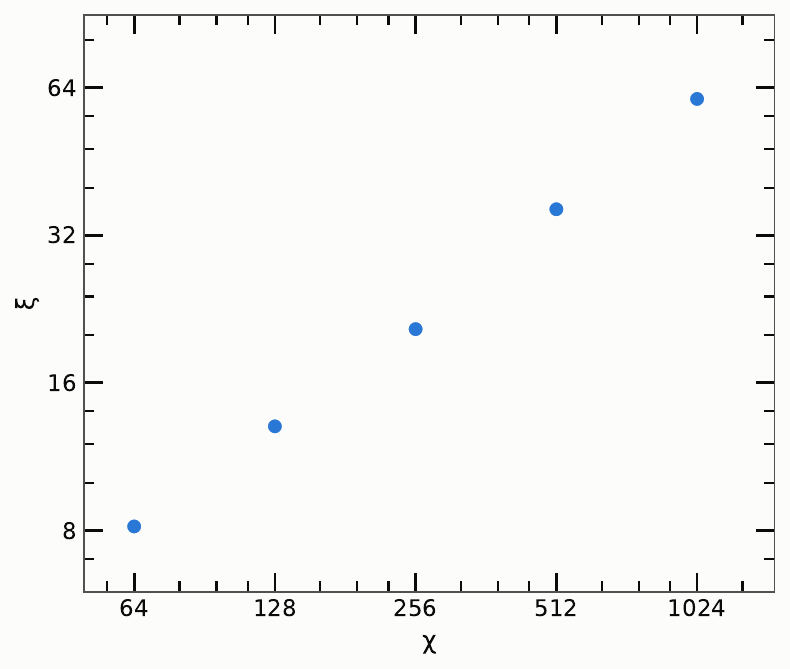}
    \caption{VUMPS correlation length estimation of trial state PEPS in Ref.~\cite{kim2016featureless}.  The correlation length without VUMPS environment truncation limitation ($\chi \rightarrow \infty$) is likely to be divergent inferring from the tendency.}
    \label{fig:fqiVUMPS}
\end{figure}

In previous studies, several works~\cite{kimchiPNAS,kim2016featureless} have focused on the question of whether there exist SU(2) and $C_{3z}$ symmetric gapped states on the honeycomb lattice without topological order.  One of the proposals is represented in PEPS form ~\cite{kim2016featureless}.  This is closely related to our investigation of possible frustrated symmetric states.  The difference is that our system does not have the mirror symmetry mapping one sublattice to the other.  Because a gapped system with no topological order is not ruled out by known LSM-type theorems, we are interested in a numerical demonstration of whether any such gapped state can adiabatically connect to the ground state in our model, which could provide counterevidence against a possible gapless spin liquid.  However, using our VUMPS algorithm rather than the old cylindrical compactification approximation with DMRG, we find the state in Ref.~\cite{kim2016featureless} is most likely to be gapless, as shown in Fig.~\ref{fig:fqiVUMPS}.  The question of the existence of a trivially gapped state in this setting, even with relaxed symmetry restrictions, thus remains unsettled.

\end{document}